\documentclass[11pt,english]{smfart}

\usepackage[T1]{fontenc}
\usepackage[english]{babel}

\usepackage{amssymb}

\usepackage{graphicx}
\usepackage{array}

\title[Effective resummation methods for an implicit resurgent  function]{
Effective  resummation  methods for an implicit
resurgent function}
 
\author{Eric Delabaere}
\address{D\'epartement de Math\'ematiques, UMR CNRS 6093,
Universit\'e d'Angers, 2 Boulevard Lavoisier, 49045 Angers Cedex 01,
France.}
\email{eric.delabaere@univ-angers.fr}
\urladdr{}


\newtheorem{Theorem}{Theorem}[section]
\newtheorem{Proposition}{Proposition}[section]
\newtheorem{Lemma}{Lemma}[section]

\theoremstyle{definition}
 
\newtheorem{Definition}{Definition}[section]
\newtheorem{Notation}{Notation}[section]

\newtheorem{rem}{Remark}[section]

\def\X{ {^\flat X}}
 
\begin{document}

\begin{abstract}
Our main aim in this self-contained article is at the same time to
detail the relationships between the resurgence and the hyperasymptotic
theories, and to demonstrate how these theories can be used  for
an implicit resurgent function.  For this purpose 
we consider after Stokes the question of the effective
Borel-resummation of an exact Bohr-Sommerfeld-like 
implicit resurgent function  whose values on
an explicit  semi-lattice provide the zeros of the Airy function. The
resurgent structure encountered resembles what one usually gets in
nonlinear problems, so that the method described here is quite general.
\end{abstract}

\dedicatory{Dedicated to C\'ecile}

\maketitle
 \section{Introduction and summary}

The problem of computing the sum of a Borel-resummable divergent 
series expansion has already a long history, which traces back to the Stokes's
epoch marking paper \cite{Sto57} where an analogous  of the
 summation to the least term (the  ``m\'ethode des
 astronomes'' of Poincar\'e \cite{Poin87}) play a key argument. 
Various methods have been developed since then, most of them being
 justified in the  framework of the Gevrey theory \cite{R93, RamSch96,
   Can99, Tho90, Bal}, see also \cite{Jen}. One of them, which will be
 used in this paper, is the resummation by factorial series
 \cite{Wn12, Nev18, Nor26,Mal95, W65} and its
 recent extension  \cite{DelabRaso06-2}.

 In most applications, the Borel resummable divergent series expansion  enjoys
 the property of being resurgent  \cite{Ec81-1, Ec81-2, Ec85, CNP2,
   DP99, D94, Costin98, Costin001, Olive, Kawai04}. 
In this case the Borel sum can be calculated by the
 hyperasymptotic theory \cite{BeH91, Old96, Old97}. The efficiency  of
 this  method has been demonstrated in various problems 
\cite{Old98, Howls03, Old05, Ho97, Delab02}.\\
These problems (mutiple integrals, linear and nonlinear ODE's,
difference equations, PDE's, ...) have a common feature : the resurgent
properties of the divergent series to be resummed, 
that is, roughly speaking, the Riemann sheet structure of its Borel
transform, is basically 
governed by a the so-called ``formal integral'' which can be derived
directly from the problem. 
This means that  the various series expansions playing a role in the
hyperasymptotics are known,  up to the Stokes
multipliers which have (and can be)  computed in the hyperasymptotic
scheme. This certainly explain why up to now no general links have
been written between resurgence and hyperasymptotic theories.

The main goal of this self-contained article is precisely to provide the 
relationships between the resurgence and hyperasymptotic theories. This
will be done by the way of an example. We shall examine 
the effective resummation of a divergent series defined implicitly, so
that no formal integral is computable. Nevertheless, as we shall see, 
the resurgent structure can be
described through general tools from resurgent theory, and this allows
hyperasymptotic expansions.

We shall be interested here in a celebrated test problem, the calculation
of  the zeros of the Airy function
\begin{equation}
Ai(k) = \frac{1}{2\pi} \int_{-\infty}^{+\infty} \cos (ks +
\frac{s^3}{3}) \, ds.
\end{equation}
In \cite{BeH91}, Berry and Howls have already shown how the
hyperasymptotics can be used to solve this problem, their method being
based on the hyperasymptotics of the Airy function itself. Our
approach differs from theirs and resembles (and by the way justify) the idea of 
Stokes \cite{Sto57} who first translates the problem into a simpler
implicit resurgent-resummable equation  and formally solve it. This
can be rigorously justified in the framework of resurgence
analysis, leading to the notion of ``model equation'' through a
resurgent-resummable change of variable (this is a key-idea in many
problems, see, e.g., \cite{DP97, DP99, DT00, Costin001}). This will
be done in \S \ref{section2}. What remains
to do then is to Borel-sum the change of variable. For that purpose we
first use, in \S 
\ref{section3}, the direct method by factorial series as explained in
\cite{DelabRaso06-2}. We then turn to the hyperasymptotics in 
\S \ref{section5}. However,  to apply this second method 
we first have to analyze the resurgent properties of our implicitly
defined formal
series. This is what we do  in  \S \ref{section4}. Using the
alien differential calculus, it can be derived that
the Riemann sheet structure of the Borel transform is analogous to
that of a solution of a nonlinear differential  equation, a Riccati equation for instance
\cite{Old05}, with a one-dimensional lattice of singularities. This
means that our method can be adapted to a large class of problems, for
instance to the so-called exact Bohr-Sommerfeld equations \cite{DDP97,
DT00, Kawai02} which are nowadays quite common in physics.

\section{The zeros of the Airy function: the Stokes-Borel approach}\label{section2}

\begin{Notation}
In this article, if $\{\omega_n\}$ is a set of points in $\mathbb{C}$
with no accumulation point, we  note:
\begin{itemize}
\item  $\mathbb{C}_{\{\omega_n\}} = \mathbb{C} \backslash
{\{\omega_n\}}$ and 
$\displaystyle 
\begin{array}{c}
\mathbb{C}_{\{\omega_n\}}^\infty \\
\pi \downarrow \\
\mathbb{C}_{\{\omega_n\}}
\end{array}$ its universal covering.
\item For $\rho >0$ small enough we write
$\displaystyle \mathbb{C}_{\{\omega_n\},\rho} = \mathbb{C}
\backslash \bigcup_{n} \overline{D(\omega_n, \rho)}$ where
$\overline{D(\omega, \rho)}$ is the close disc centered on $\omega$
with radius $\rho$, and 
$\displaystyle 
\begin{array}{c}
\mathbb{C}_{\{\omega_n\}, \rho}^\infty \\
\pi_\rho \downarrow \\
\mathbb{C}_{\{\omega_n\}, \rho}
\end{array}$ its universal covering.
\item For  $\zeta \in
\mathbb{C}_{\{\omega_n\}}^\infty$ ({\em resp.}  $\zeta \in
\mathbb{C}_{\{\omega_n\}, \rho}^\infty$) we write  $|\zeta| :=
|\pi(\zeta)|$  ({\em resp.} $|\zeta| := |\pi_\rho (\zeta)|$).
\end{itemize}
For a given formal series expansion $\displaystyle \varphi  (z) =
\sum_{n=0}^\infty \frac{\alpha_n}{z^n} \in \mathbb{C}[[z^{-1}]]$:
\begin{itemize}
\item  Its {\em minor} is defined as
 $\displaystyle  \widetilde{ \varphi } (\zeta) = \sum_{n=1}^\infty
 \alpha_n \frac{\zeta^{n-1}}{(n-1)!}$, that is its  formal Borel
 transform  when forgetting its constant term $\alpha_0$.
\item One says that $\displaystyle \varphi$ is a {\em small} formal series
expansion if $\alpha_0=0$.
\item  $\varphi$ is said to be Gevrey-1, and we note $\varphi
\in \mathbb{C}[[z^{-1}]]_1$, if its minor $\displaystyle  \widetilde{
  \varphi }$ converges at the origin.
\end{itemize}
\end{Notation}

We first recall some well-known fact on the Airy function, but this
will help us introducing necessary notations and definitions. 

\subsection{The Airy function as a Borel sum}

It is known since Stokes that the asymptotics of the Airy function is
essentially governed by the series expansion
\begin{equation}\label{varphiai}
\varphi_{Ai}  (z) =   \sum_{n=0}^\infty \frac{a_n}{z^n},
\hspace{5mm} \mbox{with} \hspace{5mm}  a_n = (-\frac{3}{4})^n
\frac{\Gamma (n+1/6)\Gamma (n+5/6)}{2\pi \Gamma (n+1)}.
\end{equation}
This series expansion is divergent but enjoys the following
properties (see, e.g., \cite{Jidoumou}):

\begin{Proposition}\label{propvarphiai}
The series expansion $\varphi_{Ai}  (z)$ is Gevrey-1 and its minor 
 $\displaystyle \widetilde{ \varphi_{Ai}} (\zeta) \in
 \mathbb{C}\{\zeta\}$ 
 extends analytically to 
$\mathbb{C}_{\{0, -4/3\}}^\infty$. Moreover, for any $\rho >0$ and
$B>0$, there exists $A=A(\rho, B)$ such that 
$$\forall \zeta \in
\mathbb{C}_{\{0, -4/3\}, \rho}^\infty, \, \, |\widetilde{
  \varphi_{Ai}} (\zeta)| \leq Ae^{B|\zeta|}.$$
\end{Proposition}

This proposition implies that for any $\displaystyle \theta \in 
\frac{\mathbb{R}}{2\pi \mathbb{Z}} \backslash \{ \pi \}$, the function
\begin{equation}\label{somborelai}
\displaystyle
\mbox{\sc  s}_{\theta} \varphi_{Ai}  (z)  =
a_0+\int_0^{\infty e^{i\theta}} \widetilde{ \varphi_{Ai}} (\zeta)  e^{-z \zeta} \, d\zeta.
\end{equation}
is a well-defined holomorphic function in $\Re(ze^{i\theta})> 0$:
$\varphi_{Ai}$ is {\em Borel-resummable  in the direction
  $\theta$} and $\mbox{\sc  s}_{\theta} \varphi_{Ai}$ is its {\em Borel-sum}
in that direction  whose asymptotics is given by $\varphi_{Ai}$: 
 Proposition \ref{propvarphiai}  and a theorem of
Nevanlinna \cite{DelabRaso06-2, Mal95} induce for instance that, for
any $\displaystyle \theta \in 
\frac{\mathbb{R}}{2\pi \mathbb{Z}} \backslash \{ \pi \}$, there exists
$r >0$ such that, for any $B>0$, 
$$\exists A >0, \, \,
\forall \, \Re(ze^{i\theta})> B, \, \, \forall \, n \geq 0, \, \,  
  \Big|\mbox{\sc  s}_{\theta} \varphi_{Ai}  (z) -\sum_{k=0}^{n}
  \frac{a_k}{z^k}\Big| \leq  Ae^{Br}  \frac{ n!  }{r^n} \frac{1}{|z|^{n}(\Re
    (ze^{i\theta})-B)}.
$$
The link between $\varphi_{Ai}$ and the Airy function is given by
the following proposition \cite{Jidoumou}:

\begin{Proposition}\label{linkAiai}
For $|\arg (z)| < \pi/2$, $|z|>0$, {\em resp.} $|\arg (k)| < \pi/3$, $|k|>0$, 
$$\displaystyle 2\sqrt{\pi} k^{1/4} Ai(k) =
 e^{-2z/3} \mbox{\sc  s}_{0}
 \varphi_{Ai}  (z) \hspace{5mm} \mbox{where}  \hspace{5mm} k= z^{2/3}. 
$$
\end{Proposition}
(Here and in the sequel we use the convention that $t^\alpha =
|t|e^{i\alpha \arg(t)}$).\\
To analytically continue $\mbox{\sc  s}_{0}  \varphi_{Ai}  (z)$ one
just has, by Cauchy, to rotate the direction of summation $\theta$ in
(\ref{somborelai}). For $\theta \in [0, -\pi[$ one thus gets: \\
for $|\arg (z) + \theta| < \pi/2$, $|z|>0$, {\em resp.} 
$|\arg (k) + 2\theta/3| < \pi/3$, $|k|>0$, 
$$\displaystyle 2\sqrt{\pi} k^{1/4} Ai(k) =
 e^{-2z/3} \mbox{\sc  s}_{\theta}
 \varphi_{Ai}  (z). 
$$

\begin{figure}[thp]
\begin{center}
\includegraphics[width=3.0in]{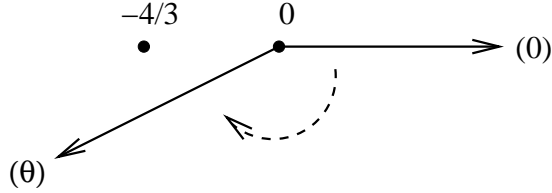} 
\caption{
Rotating the direction of Borel-resummation.
\label{fig:figart1}}
\end{center}
\end{figure}

\begin{figure}[thp]
\begin{center}
\begin{tabular}{ccc}
\includegraphics[width=2.0in]{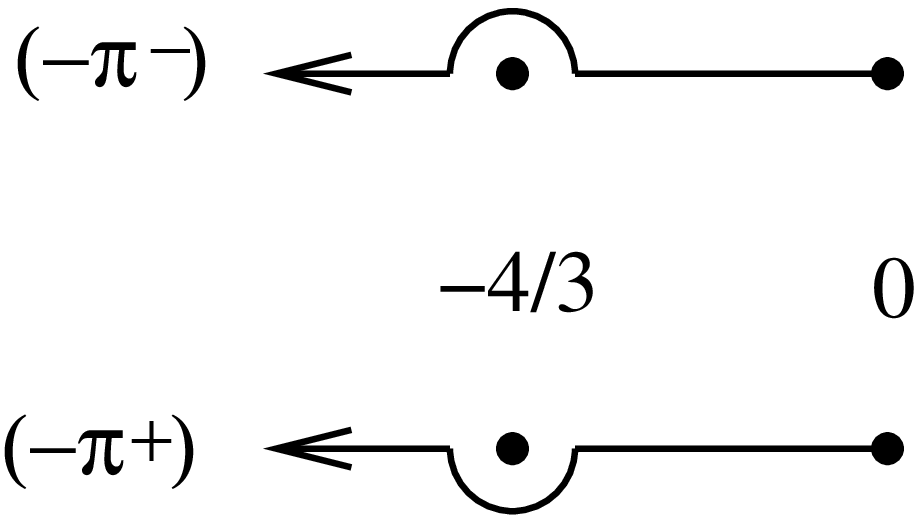} & \hspace{15mm} &
\includegraphics[width=2.0in]{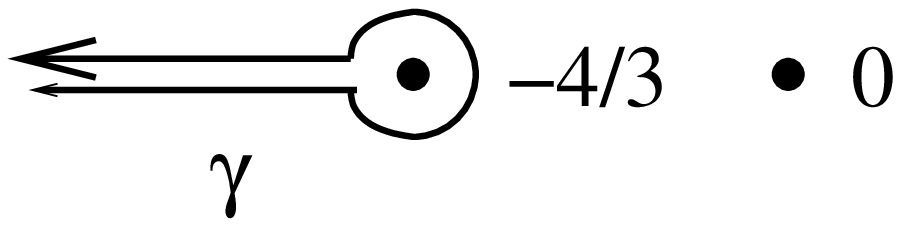} 
\end{tabular}
\caption{Right and left Borel-resummation.
\label{fig:figart2}}
\end{center}
\end{figure}

A Stokes phenomenon occurs for the $-\pi$ direction, since in that
direction one meets a singularity for $\widetilde{ \varphi_{Ai}}$ : $ \varphi_{Ai} $
is not Borel-resummable in that direction, but {\em right and left
  Borel-resummable}. For the right-resummation 
$$\mbox{\sc  s}_{-\pi^+}  \varphi_{Ai}  (z) = 
a_0+\int_0^{\infty e^{-i\pi^+}} \widetilde{ \varphi_{Ai}} (\zeta)
e^{-z \zeta} \, d\zeta.
$$
one integrates along a path avoiding the singularity as shown on
Fig. \ref{fig:figart2}. The left-resummation $\mbox{\sc  s}_{-\pi^-}
\varphi_{Ai}  (z)$ is defined in a similar way.

One can compares right and left-resummations, since 
\begin{equation}\label{leftright}
\mbox{\sc  s}_{-\pi^-}  \varphi_{Ai}  (z)  = \mbox{\sc  s}_{-\pi^+}
\varphi_{Ai}  (z)  + \int_\gamma \widetilde{ \varphi_{Ai}} (\zeta)
e^{-z \zeta} \, d\zeta
\end{equation}
where the path $\gamma$ is drawn on Fig. \ref{fig:figart2}. It can be
shown  \cite{Jidoumou} that, locally near $\zeta = -4/3$, $\widetilde{
  \varphi_{Ai}} (\zeta)$ reads
\begin{equation}\label{singai1}
\widetilde{ \varphi_{Ai}} (\zeta) = \frac{b_0}{2i\pi (\zeta +4/3)} +
\widetilde{ h}   (\zeta+4/3)  \frac{\ln(\zeta +4/3)}{2i\pi} + hol(\zeta +4/3)
\end{equation}
where $\widetilde{ h}$, $hol$ are holomorphic functions near
$0$. 

\begin{figure}[thp]
\begin{center}
\includegraphics[width=1.5in]{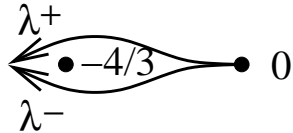} 
\caption{
The paths of analytic continuations $\lambda^{\pm}$.
\label{fig:figart4}}
\end{center}
\end{figure}
Note that $b_0$ is the residue at $-4/3$ of the analytic continuation of
$ \widetilde{ \varphi_{Ai}}$, while the $\widetilde{ h}$ may be defined as
$$\widetilde{ h} (\zeta+4/3)  = \lambda^+ \widetilde{ \varphi_{Ai}}(\zeta) - 
\lambda^- \widetilde{ \varphi_{Ai}}(\zeta)$$
where $\lambda^{\pm}$ are the paths of analytic continuation in the
direction $(-\pi)$ drawn on Fig. \ref{fig:figart4}.\\
Since $\widetilde{ h}$ is  holomorphic near the origin, it can be considered
as the minor of a (Gevrey-1) series expansion $h \in
\mathbb{C}[[z^{-1}]]$. Fixing its constant term to be $b_0$, we translate
these informations  into a single formula, namely
$$\Delta^z_{-4/3} \varphi_{Ai} (z)= h(z).$$
The operator $\Delta^z_{-4/3}$ is the so-called {\em alien derivation}
at $-4/3$ (the superscript $z$ is just added here to remind the name
of the variable). The following proposition precise 
the effect of the action of $\Delta^z_{-4/3}$ on $ \varphi_{Ai}$ \cite{Jidoumou}:

\begin{Proposition}\label{Resuai}
One has
$$\Delta^z_{-4/3} \varphi_{Ai} = -i \varphi_{Bi} \hspace{5mm}
\mbox{where} \hspace{5mm} 
\varphi_{Bi}  (z) =   \sum_{n=0}^\infty (-1)^n \frac{a_n}{z^n} = \varphi_{Ai}  (-z).$$
\end{Proposition}
Note that $\varphi_{Bi}  (z) =\varphi_{Ai}  (-z)$ implies that 
$\widetilde{ \varphi_{Bi}} (\zeta) = -\widetilde{ \varphi_{Ai}}
(-\zeta)$. From propositions \ref{propvarphiai} and \ref{Resuai} one
easily gets:

\begin{Proposition}\label{propvarphibi}
The series expansion $\varphi_{Bi}  $ is Gevrey-1 and its minor  
 $\displaystyle \widetilde{ \varphi_{Bi}}$  extends analytically to 
$\mathbb{C}_{\{0, +4/3\}}^\infty$. For any $\rho >0$ and
$B>0$, there exists $A=A(\rho, B)$ such that 
$$\forall \zeta \in
\mathbb{C}_{\{0, 4/3\}, \rho}^\infty, \, \, |\widetilde{
  \varphi_{Ai}} (\zeta)| \leq Ae^{B|\zeta|}.$$
Moreover,
$$
\Delta^z_{4/3} \varphi_{Bi} = -i \varphi_{Ai}.$$
\end{Proposition}

\begin{rem}
Propositions \ref{propvarphiai}, \ref{Resuai} and \ref{propvarphibi}
imply that $\varphi_{Ai}$ and $\varphi_{Bi}$ belong to the algebra  RES
of simply ramified resurgent functions (see \cite{GS001}).
\end{rem}

Now a direct consequence of (\ref{leftright}) is that 
$$\int_\gamma \widetilde{ \varphi_{Ai}} (\zeta)
e^{-z \zeta} \, d\zeta = e^{+4z/3}\left(b_0+ 
\int_0^{\infty e^{-i\pi}} \widetilde{ h} (\zeta)
e^{-z \zeta} \, d\zeta
 \right) = e^{+4z/3} \mbox{\sc  s}_{-\pi} \big( \Delta^z_{-4/3}
 \varphi_{Ai} \big)(z).$$
Returning to the Airy function, what we have got so far is that, 
for $|\arg (z) -\pi| < \pi/2$, $|z|>0$, {\em resp.} 
$|\arg (k) - 2\pi/3| < \pi/3$, $|k|>0$, 
$$
\begin{array}{ll}
\displaystyle 2\sqrt{\pi} k^{1/4} Ai(k) &  = e^{-2z/3} \mbox{\sc  s}_{-\pi^+}
\varphi_{Ai}  (z) \\
 & \\
\displaystyle  & = e^{-2z/3}  \mbox{\sc  s}_{-\pi^-} \varphi_{Ai}  (z)  
-  e^{+2z/3} \mbox{\sc  s}_{-\pi} \big( \Delta^z_{-4/3}
 \varphi_{Ai} \big)(z) \\
 & \\
\displaystyle  & = e^{-2z/3}  \mbox{\sc  s}_{-\pi^-} \varphi_{Ai}  (z)  
+ i e^{+2z/3} \mbox{\sc  s}_{-\pi} \varphi_{Bi}  (z).
\end{array}
$$
The Stokes phenomenon being analyzed, one can going on rotating the
direction of Borel-resummation. We finally arrive to the following
result:

\begin{Lemma}\label{lemme1}
For $|\arg (z) -3\pi/2| < \pi/2$, $|z|>0$, {\em resp.} 
$|\arg (k) - \pi/| < \pi/3$, $|k|>0$, 
\begin{equation}\label{eqlemme1}
2\sqrt{\pi} k^{1/4} Ai(k) = e^{-2z/3}  \mbox{\sc  s}_{-3\pi/2 } \varphi_{Ai}  (z)  
+ i e^{+2z/3} \mbox{\sc  s}_{-3\pi/2} \varphi_{Bi}  (z).
\end{equation}
\end{Lemma}

We are now in position to analyze the Airy function in a sector of the complex
plan bissected by the negative real axis, where the zeros we are
looking for lay. For this purpose  we are going to reduce our problem to a model
equation, in the spirit of Stokes \cite{Sto57}.

\subsection{The zeros of the Airy function}

One sees from lemma \ref{lemme1}  that, for $|\arg (k) - \pi/| < \pi/3$,
$|k|>0$, the zeros of the Airy function are those of the function
given by the right-hand
side of the  equality (\ref{eqlemme1}).

We introduce
\begin{equation}\label{newvar}
\psi_{Ai} (x) = \varphi_{Ai} (z),  \hspace{5mm} 
\psi_{Bi} (x) = \varphi_{Bi} (z), \hspace{5mm} \mbox{where}
\hspace{5mm}  z= xe^{3i\pi/2}. 
\end{equation}
For the resurgence  and Borel-resummation viewpoint, the change of
variable  $z = xe^{3i\pi/2}$ is quite innocent. In effect, a
Borel-resummation in $z$ in the direction  $-3\pi/2$ is transformed into
a Borel-resummation in $x$ in the direction  $0$, meanwhile
propositions \ref{propvarphiai}, \ref{Resuai} and \ref{propvarphibi}
translate into:

\begin{Proposition}\label{proppsi}
The series expansion 
 $\displaystyle \psi_{Ai} (x) = \sum_{n=0}^\infty (i)^n 
\frac{a_n}{x^n} $ ({\em resp.}  $\displaystyle \psi_{Bi} (x) = \sum_{n=0}^\infty (-i)^n 
\frac{a_n}{x^n} $)  is Gevrey-1 and its minor 
$\displaystyle \widetilde{ \psi_{Ai}}(\xi)$ ({\em resp.}  
$\displaystyle \widetilde{ \psi_{Bi}}(\xi) $) extends analytically to 
$\mathbb{C}_{\{0, +\frac{4i}{3}\}}^\infty$ ({\em resp.} $\mathbb{C}_{\{0,
  -\frac{4i}{3}\}}^\infty$). 
For any $\rho >0$ and
$B>0$, there exists $A=A(\rho, B)$ such that 
$$\forall \xi \in
\mathbb{C}_{\{0, \frac{4i}{3}\}, \rho}^\infty, \, \, |\widetilde{
  \psi_{Ai}} (\xi)| \leq Ae^{B|\xi|} \hspace{5mm} \mbox{resp.}
\hspace{5mm} \forall \xi \in
\mathbb{C}_{\{0, -\frac{4i}{3}\}, \rho}^\infty, \, \, |\widetilde{
  \psi_{Bi}} (\xi)| \leq Ae^{B|\xi|}.$$
Moreover,
\begin{equation}\label{equpsiaibi}
\Delta^x_{\frac{4i}{3}} \psi_{Ai} = -i \psi_{Bi}, \hspace{7mm}
\Delta^x_{-\frac{4i}{3}} \psi_{Bi} = -i \psi_{Ai}.
\end{equation}
\end{Proposition}
The right-hand side of the equality (\ref{eqlemme1}) becomes the function
$$
G(x)= e^{\frac{2}{3} ix }  \mbox{\sc  s}_{0} \psi_{Ai} (x) +  
ie^{-\frac{2}{3} ix} \mbox{\sc  s}_{0}\psi_{Bi} (x), \hspace{5mm} 
|\arg (x)| < \pi/2, \, |x|>0.
$$
Using the linearity of the Borel-resummation, one can write this
function as follows:
\begin{equation}\label{Mod1}
\displaystyle G(x) = 
\cos (\frac{2}{3}x - \frac{\pi}{4}) R(x) 
+ \sin (\frac{2}{3}x - \frac{\pi}{4})S(x)
\end{equation}
with 
\begin{equation}\label{Mod2}
\begin{array}{c}
R(x) = \mbox{\sc  s}_{0} r(x) , \hspace{5mm}
S(x) = \mbox{\sc  s}_{0} s(x)\\
\\
\displaystyle r(x) =  e^{i\pi /4}\psi_{Ai}  (x) + e^{i\pi /4}
\psi_{Bi}  (x) = 2e^{i\pi/4} \sum_{m=0}^\infty (-1)^m
\frac{a_{2m}}{x^{2m}} \\
\\
\displaystyle s(x) = e^{3i\pi /4}\psi_{Ai} (x) - e^{3i\pi/4} \psi_{Bi}  (x) = 
2e^{i\pi/4} \sum_{m=1}^\infty (-1)^{m}
  \frac{a_{2m-1}}{x^{2m-1}}.
\end{array}
\end{equation}

To proceed it will be convenient to use the following definition
throughout the rest of this article.

\begin{Definition}\label{defres}
We shall say that a formal series expansion $\varphi(x) \in
\mathbb{C}[[x^{-1}]]$ is resurgent if $\varphi$ is Gevrey-1 and if its
minor extends
analytically on  $\mathbb{C}_{\{\frac{4i}{3}
  \mathbb{Z}\}}^\infty$. 
\end{Definition}

In fact, Definition \ref{defres} just defines  a special class of
resurgent functions, see \cite{Ec81-1, Ec81-2, Ec85, CNP2}.

Note that, from  proposition \ref{proppsi}, the formal series
expansions $\psi_{Ai}$, $\psi_{Bi}$, and their linear combinations 
$r$ and $s$ are resurgent, and Borel-resummable in any direction
$\displaystyle \theta \in \frac{\mathbb{R}}{2\pi \mathbb{Z}}
\backslash \{\pm \pi/2 \}$.

We shall use the following theorem:

\begin{Theorem}\label{compose}
We assume that $\varphi_1, \, \varphi_2  \in \mathbb{C}[[x^{-1}]]_1$
are resurgent formal series expansions. 
\begin{enumerate}
\item The product $\varphi_1 . \varphi_2$ is a resurgent formal series
  expansion. If $\varphi_1, \, \varphi_2$  are
 Borel-resummable in a direction $\theta \in
\frac{\mathbb{R}}{2\pi \mathbb{Z}}$, then the product $\varphi_1 . \varphi_2$  is
Borel-resummable in that direction and 
$\displaystyle \mbox{\sc  s}_{\theta} \big( \varphi_1 . \varphi_2
\big)(x) = 
\mbox{\sc  s}_{\theta}  \varphi_1(x).
\mbox{\sc  s}_{\theta}  \varphi_2 (x)
$.
\item If  $\varphi_1$ is small and if 
$\displaystyle 
\Psi (\varepsilon) = \sum_{n=0}^\infty \alpha_n \varepsilon_1^n \in
\mathbb{C}\{ \varepsilon \} $ is a holomorphic function near
$\varepsilon =0$, then the composition 
$\displaystyle \Psi \circ \varphi_1 (x) = \sum_{n=0}^\infty \alpha_n
\varphi^n(x)$
is a resurgent formal series  expansion.
Moreover, if $\varphi_1$ is Borel-resummable in a direction $\theta \in
\frac{\mathbb{R}}{2\pi \mathbb{Z}}$, then $\Psi \circ \varphi_1$ is
Borel-resummable in that direction and 
$\displaystyle \mbox{\sc  s}_{\theta} \big( \Psi \circ \varphi_1 \big)(x) = \Psi \circ \big(
\mbox{\sc  s}_{\theta}  \varphi_1 \big) (x)$.
\item If $\varphi_1$ is small and if $\varphi(x) = x + \varphi_1(x)$,
  then  the composition 
$\varphi_2 \circ \varphi$ defined by 
$$\varphi_2 \circ \varphi (x)  = \sum_{n=0}^\infty \frac{\varphi_1^n(x)}{n!} 
 \frac{d^n \varphi_2 }{dx^n}(x) $$
is resurgent.  If $\varphi_1, \, \varphi_2$  are
 Borel-resummable in a direction $\theta \in
\frac{\mathbb{R}}{2\pi \mathbb{Z}}$, then $\varphi_2 \circ \varphi$ is 
Borel-resummable in that direction and 
$\displaystyle \mbox{\sc  s}_{\theta} \big( \varphi_2 \circ \varphi
\big)(x) = 
\mbox{\sc  s}_{\theta}  \varphi_2  \Big(x+
\mbox{\sc  s}_{\theta}  \varphi_1  (x) \Big)
$.
\end{enumerate}
\end{Theorem}
Apart from the Borel-resummability, this theorem is just a specialization
of more general theorems in resurgence theory \cite{CNP2}. In
our case the proof can be done in  a simpler way  by the methods used in \cite{GS001,
  DelabRaso06-1, Tou}, including the Borel-resummabilty properties
(which are also a consequence of the theorem of Ramis-Sibuya \cite{Mal95}).

This theorem \ref{compose} will allow us to write (as Stokes do \cite{Sto57}) the
functions $R$ and $S$ in a polar form:
\begin{equation}\label{polar}
R(x)=M(x) \cos \big(\Phi(x)\big) \hspace{10mm} S(x)= M(x) \sin
\big(\Phi(x)\big).
\end{equation}
One first defines $M(x)$. Theorem
\ref{compose} implies that the formal  series
expansion 
$$m(x) = (r^2(x) + s^2(x))^{1/2} = 2 e^{i\pi /4}  \psi^{1/2}_{Ai} (x)
\psi^{1/2}_{Bi} (x)$$
is resurgent and  Borel-resummable in any direction
$\displaystyle \theta \in \frac{\mathbb{R}}{2\pi \mathbb{Z}}
\backslash \{\pm \pi/2 \}$.  We define $M(x)$ as the Borel-sum of
$m(x)$ in the direction $0$. \\
Note that, since $m$ is not small, it is {\em invertible} (this is
again a consequence of theorem \ref{compose}). This ensures that, for
$\Re (x)$ large enough,  $M(x)$ does not vanish.

We no define $\Phi(x)$: $r$ being invertible and $s$ being small,
$s/r$ is a small resurgent formal series expansion, Borel-resummable 
in any direction
$\displaystyle \theta \in \frac{\mathbb{R}}{2\pi \mathbb{Z}}
\backslash \{\pm \pi/2 \}$. 
Applying again theorem  \ref{compose}, the same properties will be
true for the formal series expansion $\displaystyle  
\phi(x) =  \arctan \big(   \frac{s(x)}{r(x)} \big)$, and one can define
\begin{equation}\label{defPhi}
\Phi(x) = \arctan \frac{S(x)}{R(x)} 
 = \arctan  \circ \, \mbox{\sc  s}_{0} \Big(   \frac{s}{r} \Big)(x) = 
\mbox{\sc  s}_{0} \Big( \arctan  \circ    \frac{s}{r} \Big)(x).
\end{equation}
The polar form (\ref{polar}) being justified, one deduces from
(\ref{Mod1}) that:
$$ G(x) = M(x) \cos \big(\frac{2}{3}x - \frac{\pi}{4} -
\Phi (x) \big).$$
We summarize what we have obtained:

\begin{Theorem}\label{theomp2}
For $\Re (x)$ large enough, the zeros of the Airy function $Ai(k)$
with $k = e^{i\pi}x^{2/3}$  are
the solutions of the equation
\begin{equation}\label{presmodel}
\cos \big(\frac{2}{3}x - \frac{\pi}{4} -
\Phi (x) \big)=0,
\end{equation}
with  $\Phi (x) = \mbox{\sc  s}_{0} \phi(x)$, where $\displaystyle  
\phi(x) =  \arctan \big(   \frac{s(x)}{r(x)} \big)$ is a
 small resurgent formal series expansion, Borel-resummable 
in any direction
$\displaystyle \theta \in \frac{\mathbb{R}}{2\pi \mathbb{Z}}
\backslash \{\pm \pi/2 \}$.
\end{Theorem}

\subsection{The model equation}

The  {\em model equation} for our problem is the equation
\begin{equation}\label{modeleq}
\cos(\frac{2}{3}t -\frac{\pi}{4})  = 0, \hspace{5mm} \Re (t) >0,
\end{equation}
deduced from (\ref{presmodel}) through the change of variable 
\begin{equation}\label{vart}
t = x -  \frac{3}{2}\Phi(x).
\end{equation}
The model equation (\ref{modeleq}) is obviously exactly solvable, the solutions being
\begin{equation}\label{sol1}
t = \frac{3}{2}(l-\frac{1}{4})\pi, \hspace{5mm} l \in  \mathbb{N}^\star.
\end{equation}
To translate this result in term of the zeros of the Airy function,
one has to justify the change of variable (\ref{vart}) and to
calculate the inverse function. \\
We know that $\Phi(x) = \mbox{\sc  s}_{0} \phi (x)$ where
$\displaystyle \phi(x)$ is a small resurgent formal
series expansion. Let us look at (\ref{vart}) at a formal level, that
is
$$t = x -  \frac{3}{2}\phi(x), \hspace{5mm} \mbox{or equivalently}
\hspace{5mm}  x = t +  \frac{3}{2}\phi(x).
$$
This fixed point problem has a unique formal solution $X(t)$, $X(t)-t
\in \mathbb{C}[[t^{-1}]]$, which
 can be constructed  by the formal successive approximation method: if
\begin{equation}\label{iterX}
\left\{
\begin{array}{l}
\displaystyle X_0(t) = t\\
\displaystyle X_{n+1} (t) = t + \frac{3}{2}\phi \circ X_{n} (t), \, \,
\, n \geq 0
\end{array}
\right.
\end{equation}
then one easily checks that the sequence $(\X_n)=
(X_n -t)$ converges in the algebra of formal series expansions 
$\mathbb{C}[[t^{-1}]]$ to a unique small series expansion $f \in
\mathbb{C}[[t^{-1}]]$,
 since for all  $n \in \mathbb{N}$,  
$X_{n+1} - X_n \in t^{-n-1}\mathbb{C}[[t^{-1}]]$.
Note that, from theorem \ref{compose}, every $\X_n$ is a small
resurgent formal series expansion, and  Borel-resummable in any direction
$\displaystyle \theta \in \frac{\mathbb{R}}{2\pi \mathbb{Z}}
\backslash \{\pm \pi/2 \}$ (by iteration, since this is true for
$\phi$).  Applying the resurgent implicit function theorem
\cite{CNP2}, one shows that the limit $\X(t)$ is a small
resurgent formal series expansions, Borel-resummable in any direction
$\displaystyle \theta \in \frac{\mathbb{R}}{2\pi \mathbb{Z}}
\backslash \{\pm \pi/2 \}$ (this can be obtained by direct estimates
or by the Ramis-Sibuya theorem). Also, by construction, writing $X(t)
= t + \X(t)$:

\begin{Proposition}
There exist two sectorial neighbourhood of infinity $\Sigma_x$ and
$\Sigma_t$ of aperture
$I=]-\pi/2, \pi/2[$ such that
$$x \in \Sigma_x, \, \, t =  x -  \frac{3}{2}\Phi(x) \Leftrightarrow t
\in \Sigma_t, \, \, x= \mbox{\sc  s}_{0} X(t).$$
\end{Proposition}

In this proposition  $\mbox{\sc  s}_{0} X(t) = t + \mbox{\sc  s}_{0}
\X(t)$ and:

\begin{Definition}
A  sectorial neighbourhood of infinity of aperture $]-\pi/2, \pi/2[$
 is an open set $\Sigma$ of $\mathbb{C}$ 
such that for any open interval $J \subset I$, there
is $z \in \Sigma$ such that $zJ \subset \Sigma$, where 
$\displaystyle zJ :=
\{z+re^{i\theta}, \, r>0, \, \theta \in J\}$.
\end{Definition}

Returning to theorem (\ref{theomp2}) what we have obtained is
summarized in the following theorem.

\begin{Theorem}\label{theomp1}
For $\Re (x)$ large enough, the zeros of the Airy function $Ai(k)$
with $k = e^{i\pi}x^{2/3}$  are given by
\begin{equation}\label{zero1}
x_l = \mbox{\sc  s}_{0} X \big( \frac{3}{2}(l-\frac{1}{4})\pi \big), 
\hspace{5mm} l \in  \mathbb{N}^\star, \,\, l \mbox{ large enough}.
\end{equation}
where $X(t)$ is the unique formal solution of the implicit equation
\begin{equation}\label{iterX2}
 X (t) = t + \frac{3}{2}\phi \circ X (t), \hspace{5mm} \mbox{with}
 \hspace{5mm}
\phi(x) =  \arctan \big(   \frac{s(x)}{r(x)} \big).
\end{equation}
This formal solution $X(t)$ reads $X(t) = t + \X(t)$ where $\X(t)$ is 
a small resurgent formal
series expansion, Borel-resummable  in any direction
$\displaystyle \theta \in \frac{\mathbb{R}}{2\pi \mathbb{Z}}
\backslash \{\pm \pi/2 \}$.
\end{Theorem}

\section{The zeros of the Airy function: calculation with the
  factorial series method}\label{section3}

To compute the zeros of the Airy function, we first have now to
calculate the series expansion $X(t)$. Following theorem
\ref{theomp1}, this first means to expand $\phi(x)$ which is
straightforward. We then apply the formal successive approximation
method (\ref{iterX}). With Maple V Release 5.1 one gets the result to
any fixed order:
\begin{equation}\label{expXt}
X(t) = t + \frac{5}{32}t^{-1} - \frac{1255}{6144}t^{-3}
+ \frac{272075}{196608}t^{-5} + \cdots =
t + \sum_{n=0}^\infty \frac{c_n}{t^n}
\end{equation}
As we shall see, this sole information on $X(t)$ is quite enough to
calculate the zeros of the Airy function to any order, using the
factorial series method.

\begin{figure}[thp]
\begin{center}
\includegraphics[width=2.5in]{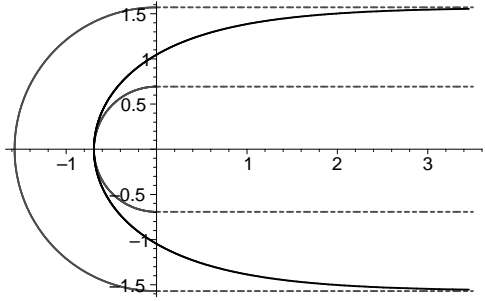} 
\caption{The open sets  $\displaystyle  \mathcal{B}_{\ln(2)}
\subset \Delta  \subset \mathcal{B}_{\frac{\pi}{2}}$.
\label{fig:Som_Factorielle06}}
\end{center}
\end{figure}

We introduce some  notations:

\begin{Notation}\label{Notation2}
\begin{itemize}
\item  For $r>0$ we note 
$\displaystyle \mathcal{B}_r  =\{\tau  \in
    \mathbb{C} \, /  \, d(\tau, \mathbb{R}^+) < r\}$, 
where $d$  is the euclidian distance measure.\\
\item We note  $\Delta$ the image of the open disc $D(1,1)$ centered
  on $1$ with radius $1$ under the biholomorphic mapping
$\displaystyle s \in D(1,1) \mapsto \tau = -\ln (s) \in \Delta$.\\ 
The open set $\Delta $ satisfy :
$$\displaystyle  \mathcal{B}_{\ln(2)}
\subset \Delta  \subset \mathcal{B}_{\frac{\pi}{2}} \hspace{5mm} 
\mbox{(cf. Fig. \ref{fig:Som_Factorielle06})}.$$
\item For  $\lambda >0$, $\Delta_\lambda$ is the homothetic set of 
$\Delta$ defined by: 
$\displaystyle \Delta_\lambda = \{\lambda \tau \, /  \, \tau \in  \Delta \}$.
\end{itemize}
\end{Notation}

We shall use the following theorem whose proof is detailed in \cite{DelabRaso06-2}:

\begin{Theorem}\label{thmsomfact1ter}
Assume that the minor 
$\displaystyle \widetilde{h}(\tau)$ of the formal series expansion
$\displaystyle h(t)=\sum_{n=0}^{+\infty} \frac{\alpha_n}{t^n} \in
\mathbb{C}[[z^{-1}]]_1$ extends analytically to the open set
$\Delta_\lambda$, $\lambda>0$. Assume furthermore that there exist $A>0$ and
$B>0$ such that for every  $\tau \in \Delta_\lambda$,
$\displaystyle |\widetilde{f}(\tau)|\leq Ae^{B|\tau|}$. Then: \\
$\bullet$  the factorial series  $\displaystyle 
\alpha_0  +\lambda \sum_{n=0}^{N}
\frac{\Gamma(\lambda t) \Gamma(n+1) \beta_n^{(\lambda)}}{\Gamma(\lambda
  t+n+1)}$ converges  absolutly for   $\Re (t) > \max( B, 1/\lambda)$,
its sum being the Borel-sum
$\mbox{\sc  s}_{0} h(t)$ of $h$, the
$\beta_n^{(\lambda)}$ being deduced from the  $\alpha_n^{(\lambda)} = \lambda^{n-1} \alpha_n$ 
by the Stirling algorithm (proposition \ref{algoStirling}).\\
$\bullet$ For any $N \geq 0$ and $\Re (t) > B $, 
\begin{equation}\label{remarquablelamb}
\begin{array}{c}
\displaystyle 
 \left| \mbox{\sc  s}_{0} h (z) - \Big( \alpha_0  +\lambda \sum_{n=0}^{N}
\frac{\Gamma(\lambda t) \Gamma(n+1) \beta_n^{(\lambda)}}{\Gamma(\lambda t+n+1)}
\Big)\right| \leq  R_{fact}(\lambda, A,B,N,t) \\
\\
\displaystyle R_{fact}(\lambda, A,B,N,t) = \hspace{80mm} \, \\
\displaystyle  \,\hspace{20mm}  \frac{A}{(\lambda B)^{\lambda B}} \frac{\, (N+\lambda
   B+1)^{N+\lambda B+1}
  }{ (N+1)^{N} } 
\left| \frac{\Gamma (\lambda t) \Gamma (N+1)}{\Gamma(\lambda t+N+1) (\Re(t) -B)}
\right|,
\end{array}
\end{equation}
\end{Theorem}

\begin{Proposition}[Stirling algorithm]\label{algoStirling}
$$ \displaystyle \forall n \geq 0, \quad \beta_n=\frac{1}{n
  !}\sum_{k=1}^{n+1} (-1)^{n-k+1} \mathfrak{s}(n,k-1) \alpha_k,$$ where
$\mathfrak{s}(n,k)$ are the Stirling cycle numbers (or Stirling numbers of the first kind):
$\displaystyle \prod_{k=0}^{n-1} (x-k) = \sum_{k=0}^n \mathfrak{s} (n,k) x^k$.
\end{Proposition}

From theorem \ref{theomp1} we see that theorem  \ref{thmsomfact1ter}
can be applied to the formal series expansion $\X(t)$ as soon as one
chooses $\lambda \in ]0, 4/\pi[$, so that the open set
$\Delta_\lambda$ is included in
th cut plane   $\mathbb{C} \backslash [\pm i4/3,
\pm i\infty[$. Therefore:\\
{\em  for any $\lambda \in ]0, 4/\pi[$, there exist
$A>0$ and $B>0$ such that, for $\Re(t) > B$ and $N \geq 0$,
$$ \left| \mbox{\sc  s}_{0} X(t) - \Big( t  +\lambda \sum_{n=0}^{N}
\frac{\Gamma(\lambda t) \Gamma(n+1) \beta_n^{(\lambda)}}{\Gamma(\lambda t+n+1)}
\Big)\right| \leq  R_{fact}(\lambda, A,B,N,t)$$
where the the
$\beta_n^{(\lambda)}$ are deduced from the  $c_n^{(\lambda)} =
\lambda^{n-1} c_n$ (cf. (\ref{expXt})) by the Stirling algorithm.
}
As explained in \cite{DelabRaso06-2}, in pratice for $\Re (t)$ and $N$
large enough, one can evaluate the
remainder term by:
\begin{equation}\label{erroest}
R_{fact}(\lambda, A,B,N,t) \sim | \beta_{l,N+1}^{(\lambda)}| 
 \frac{|\Gamma (\lambda t)| \Gamma (N+1)}{\Re( t)
   |\Gamma(\lambda   t+N+1)|}.
\end{equation}

Following theorem \ref{theomp1}, we now evaluate the $\displaystyle 
x_l = \mbox{\sc  s}_{0} X \big( \frac{3}{2}(l-\frac{1}{4})\pi \big)$,
$l \in \mathbb{N}^\star$ by the factorial series methods, and then
translates the results (and the error estimates) to the values 
$k_l = - x_l^{2/3}$ which give the zeros of the Airy function. 
Numerical calculations give very good results, even for $l=1$. The
comparison $Exact - Est.$ with the ``exact zeros '' have been made 
taken for granted that the AiryAiZeros function of Maple V
Release 5.1 gives the correct answer.

\begin{table}[thp]
\begin{center}
\begin{tabular}{|c|l|c|}
\hline
Value of $N$ & Estimates for $k_1$  & Real error\\
\hline
21 & -2.338107342 &  $0.68\times 10^{-7}$\\
\hline
41 & -2.3381074010  & $0.95 \times 10^{-8}$ \\
\hline
61 & -2.338107410494 & $0.34 \times 10^{-10}$ \\
\hline
\end{tabular}
\caption{Calculation for the first zero $k_1$ of the Airy function. We
  have applied the factorial series method with $\lambda = 1.2$. For
  the value of $t$ used here, the error estimates (\ref{erroest})
  do not  apply  here.
\label{table1}}
\end{center}
\end{table}

\begin{table}[thp]
\begin{center}
\begin{tabular}{|c|l|c|c|}
\hline
Value of $N$ & Estimates for $k_3$ & Error estimates & Real error\\
\hline
21 & -5.5205598280955580 & $0.13 \times 10^{-12}$ & $0.69\times 10^{-14}$\\
\hline
41 & -5.520559828095551049  & $0.26 \times 10^{-16}$ & $- 0.10\times
10^{-16}$\\
\hline
61 & -5.5205598280955510591283  & $0.73 \times 10^{-19}$ & $-0.16\times 10^{-20}$\\
\hline
\end{tabular}
\caption{Calculation for the third zero $k_3$ of the Airy function. We
  have applied the factorial series method with $\lambda = 1.2$ and
  used (\ref{erroest}) for  the error estimates.
\label{table2}}
\end{center}
\end{table}

\section{Resurgent structure for the implicit function}\label{section4}

We would like to turn to hyperasymptotics. This requires first to
determine the resurgent structure of $X(t)$. Since $X(t)$ is defined
implicitly by the equation (\ref{iterX2}), we first have to precise
the resurgent structure of the small resurgent formal series expansion 
$\displaystyle \phi(x) =  \arctan \big(   \frac{s(x)}{r(x)}
\big)$. For that purpose, we
shall freely use the alien derivations.

\subsection{Resurgent structure for $\phi(x)$}

In proposition  (\ref{proppsi}) we have seen that the resurgent
structure of $ \psi_{Ai}$ and $\psi_{Bi}$, that is the singularity
structure of their minors, was governed by the alien derivatives
(\ref{equpsiaibi}), namely
$$
\left\{
\begin{array}{l}
\Delta^x_{\frac{4i}{3}} \psi_{Ai} = -i \psi_{Bi}\\
\\
\mbox{otherwise, } \, \Delta^x_{\omega} \psi_{Ai} =0
\end{array}
\right.
 \hspace{7mm}
\left\{
\begin{array}{l}
\Delta^x_{-\frac{4i}{3}} \psi_{Bi} = -i \psi_{Ai}.\\
\\
\mbox{otherwise, } \, \Delta^x_{\omega} \psi_{Bi} =0
\end{array}
\right.
$$
The alien derivations are derivations in the algebraic sense; in
particular they are linear operators.  From the very definitions
(\ref{Mod2}) of $r$ and $s$, one easily gets:
$$
\begin{array}{c}
\displaystyle \Delta^x_{\frac{4i}{3}}  r(x) =  e^{-i\pi /4}\psi_{Bi}  (x) ,
\hspace{5mm} \Delta^x_{-\frac{4i}{3}}  r(x) =  e^{-i\pi /4}\psi_{Ai}  (x) \\
\\
\displaystyle \Delta^x_{\frac{4i}{3}} s(x) = e^{i\pi /4}\psi_{Bi} (x), 
\hspace{5mm}  \Delta^x_{-\frac{4i}{3}} s(x) = - e^{i\pi/4} \psi_{Ai}  (x) 
\end{array}
$$
Applying the Leibniz chain rule, one thus gets:
$$\Delta^x_{\frac{4i}{3}} \frac{s}{r} = \frac{r \Delta^x_{\frac{4i}{3}}s - 
s \Delta^x_{\frac{4i}{3}}r}{r^2} = \frac{2 \psi^2_{Bi}}{(\psi_{Ai} +
\psi_{Bi})^2}  $$
and
$$\Delta^r_{-\frac{4i}{3}} \frac{s}{r} =  - \frac{2 \psi^2_{Ai}}{(\psi_{Ai} +
\psi_{Bi})^2}.  $$
We now need the following general result from resurgence theory
\cite{CNP2, Ec81-1}:
\begin{Theorem}\label{Deltacompose}
If  $\varphi$ is a small resurgent formal series expansion and if 
$\displaystyle  \Psi (\varepsilon)$ is a holomorphic function near
$\varepsilon =0$, then for any $\omega \in \mathbb{C}^\star$,
$$\Delta_\omega \Psi \circ \varphi =  \Big( \Delta_\omega \varphi \Big). \Psi
^\prime \circ \varphi$$
where $^\prime$ is the usual derivation.
\end{Theorem}
Thus,  for any $\omega \in \mathbb{C}^\star$, 
$\displaystyle \Delta^x_{\omega} \phi =  \Big( \Delta^x_{\omega} \frac{s}{r} \Big). \arctan^\prime
(\frac{s}{r})$.
Applying this for $\omega = \pm \frac{4i}{3}$, one obtains:

\begin{Proposition}\label{REsforphi}
The resurgent structure of $\phi(x)$ is governed by:
\begin{equation}\label{EQaphi}
\left\{
\begin{array}{l}
\displaystyle  \Delta^x_{\frac{4i}{3}} \phi 
=\frac{1}{2} \frac{\psi_{Bi}}{ \psi_{Ai} }\\
\\
\displaystyle  \Delta^x_{-\frac{4i}{3}} \phi 
=-\frac{1}{2}\frac{\psi_{Ai}}{ \psi_{Bi} }\\
\\
\mbox{otherwise, } \, \Delta^x_{\omega} \phi =0
\end{array}
\right.
\end{equation}
\end{Proposition}

\subsection{Translation in term of analytic structure}

It is certainly time to pose and to say more about
alien derivations (in a way suited to our purpose), so as to explain
the meaning and the consequences of (\ref{EQaphi}).

\subsubsection{Alien derivations and analytic continuations}

\begin{figure}[thp]
\begin{center}
\includegraphics[width=2.8in]{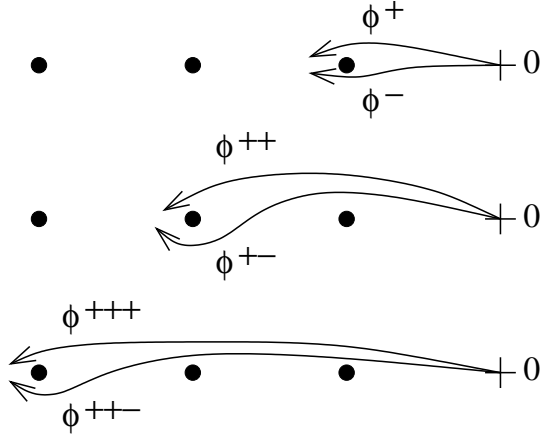} 
\caption{
The paths of analytic continuations for the direction $\pi/2$ (we have
rotated the picture to save place). The bullets are the points
$4in/3$, $n \in \mathbb{N}^\star$.
\label{fig:Exemple13}}
\end{center}
\end{figure}

Since $\phi$ is
resurgent (theorem \ref{theomp1}) its minor $\widetilde \phi$ is an
holomorphic function near $0$ which can be analytically prolonged in
the direction $\pi/2$ (say), provided to avoid the singular  semi-lattice $\frac{4i}{3}
\mathbb{N}^\star$.  \\
For $n \in \mathbb{N}^\star$ and $\xi \in \displaystyle
]\frac{4i}{3}n, \frac{4i}{3}(n+1)[$ we note 
$\displaystyle  {\widetilde \phi}^{\epsilon_1, \epsilon_2, \cdots,
  \epsilon_{n}} (\xi)$ the analytic continuation of  $\widetilde \phi
(\xi)$ along a path which circumvents each $\frac{4i}{3}k$  to the left
($\epsilon_k = +$) or to the right ($\epsilon_k = -$), see
Fig. \ref{fig:Exemple13}. 

We fix a $N \in \mathbb{N}^\star$ and consider, for $\xi \in \displaystyle
]\frac{4i}{3}(N-1), \frac{4i}{3}N[$ the mean
\begin{equation}\label{singai1bisO}
\widetilde \phi_N (\xi) = \sum_{\epsilon =(\epsilon_1, \epsilon_2, \cdots,
  \epsilon_{N-1})} \frac{p(\epsilon)! q(\epsilon)!}{N!} \, 
{\widetilde \phi}^{\epsilon_1, \epsilon_2, \cdots,
  \epsilon_{N-1}}(\xi)
\end{equation}
where the sum is made over the $2^{N-1}$ $(N-1)$-lists  
$\epsilon = (\epsilon_1, \epsilon_2, \cdots,
  \epsilon_{N-1})= (\pm, \pm \cdots, \pm)$ whereas $p(\epsilon)$ is
  the number of $+$ and  $q(\epsilon) =
  N-1-p(\epsilon)$ the number of $-$.\\
In addition to theorem \ref{theomp1}
it can be shown that the $\widetilde \phi_N$'s inherit the type of
singularities (\ref{singai1}) of $\widetilde{ \varphi_{Ai}}$ and $\widetilde{
  \varphi_{Bi}}$ , that is
\begin{equation}\label{singai1bis}
\widetilde \phi_N (\xi) = \frac{b}{2i\pi (\xi - \frac{4i}{3}N)} +
\widetilde{ h} (\xi - \frac{4i}{3}N) \frac{\ln (\xi - \frac{4i}{3}N)}{2i\pi} 
 + hol(\xi - \frac{4i}{3}N)
\end{equation}
where $\widetilde{ h}$ and $hol$ are holomorphic functions near
$0$. The coefficient $b$ is just the residue of $\widetilde \phi_N$ at
$\displaystyle \frac{4i}{3}N$ while, for  $\xi \in \displaystyle
]\frac{4i}{3}N, \frac{4i}{3}(N+1)[$,
\begin{equation}\label{DefDelta}
\widetilde{ h} (\xi - \frac{4i}{3}N)  = \sum_{\epsilon =(\epsilon_1, \epsilon_2, \cdots,
  \epsilon_{N-1})} \frac{p(\epsilon)! q(\epsilon)!}{N!} \left(
{\widetilde \phi}^{\epsilon_1, \epsilon_2, \cdots,
  \epsilon_{N-1},+}(\xi) - {\widetilde \phi}^{\epsilon_1, \epsilon_2, \cdots,
  \epsilon_{N-1},-}(\xi)\right)
\end{equation}
{\em Denoting by $h$ the inverse formal Borel transform of $\widetilde{ h}$
with constant term $b$, one gets the alien derivative of $\phi$ at 
$\displaystyle \frac{4i}{3}N$,
$$\Delta^x_{\frac{4i}{3}N} \phi (x) = h(x).$$}
The alien derivatives $\displaystyle \Delta^x_{-\frac{4i}{3}N} \phi
(x)$  are defined in a similar way. Of course 
for $\displaystyle \omega \notin \frac{4i}{3}\mathbb{Z}^\star$, 
$\displaystyle \Delta^x_\omega \phi  =0$.

From the fact that $\displaystyle  \Delta^x_{\omega} \phi =0$ when
$\displaystyle  \omega \neq \pm \frac{4i}{3}$ (proposition
\ref{REsforphi}), it would be wrong to deduce that $\widetilde \phi$
has no other ``glimpsed'' singularities in the $\pi/2$ direction than
the ``seen'' (``adjacent'' in hyperasymptotic theory) singularity 
$\displaystyle  \frac{4i}{3}$. \\
Let us consider what happens at  $\displaystyle  \frac{8i}{3}$. On the
one hand, the equality $\displaystyle 
\Delta^x_{\frac{8i}{3}} \phi  = 0$  translates into the fact that
the mean function
$$\frac{1}{2}\Big(  
{\widetilde \phi}^{+}(\xi) + {\widetilde \phi}^{-}(\xi)\Big), \hspace{5mm}
\xi \in ]\frac{4i}{3}, \frac{8i}{3}[
$$
extends holomorphically near $\displaystyle \xi = \frac{8i}{3}$. In
particular concerning the residues,
\begin{equation}\label{0stcons}
\mbox{res}_{\frac{8i}{3}}  {\widetilde \phi}^{+}(\xi) =
-\mbox{res}_{\frac{8i}{3}}   {\widetilde \phi}^{-}(\xi),
\end{equation}
whereas concerning the variations,
$$
\Big( {\widetilde \phi}^{++}(\xi) + {\widetilde \phi}^{-+}(\xi) \Big) -
\Big( {\widetilde \phi}^{+-}(\xi) + {\widetilde \phi}^{--}(\xi) \Big) =0, 
\hspace{5mm}
\xi \in ]\frac{8i}{3}, \frac{12i}{3}[
$$
which we writes as
\begin{equation}\label{1stcons}
\Big( {\widetilde \phi}^{++}(\xi) - {\widetilde \phi}^{+-}(\xi) \Big) +
\Big( {\widetilde \phi}^{-+}(\xi) - {\widetilde \phi}^{--}(\xi) \Big) =0, 
\hspace{5mm}
\xi \in ]\frac{8i}{3}, \frac{12i}{3}[.
\end{equation}
On the
other hand, from (\ref{EQaphi}), (\ref{equpsiaibi}) and the Leibniz
chain rule,
$$ \Delta^x_{\frac{4i}{3}} \Big( \Delta^x_{\frac{4i}{3}} \phi \Big) 
= \Delta^x_{\frac{4i}{3}} \Big( \frac{\psi_{Bi}}{ 2\psi_{Ai} } \Big) = \frac{i}{2}
 \frac{\psi_{Bi}^2}{ \psi_{Ai}^2 }.$$
Therefore the analytic function
$$  
{\widetilde \phi}^{+}(\xi) - {\widetilde \phi}^{-}(\xi),
$$
defined for $\displaystyle \xi \in ]\frac{4i}{3}, \frac{8i}{3}[$ (which up to a
translation in the variable space corresponds to the minor of $\displaystyle
\Delta^x_{\frac{4i}{3}} \phi$)  is singular
at $\displaystyle \xi =\frac{8i}{3}$ with a residue equal to
$\displaystyle \frac{i}{2}$
(the constant term of $\displaystyle \frac{i}{2} \frac{\psi_{Bi}^2}{
  \psi_{Ai}^2 }$) so that
\begin{equation}\label{0stconsbis}
\mbox{res}_{\frac{8i}{3}}  {\widetilde \phi}^{+}(\xi) 
-\mbox{res}_{\frac{8i}{3}}   {\widetilde \phi}^{-}(\xi) =\frac{i}{2} ,
\end{equation}
 while
$$ 
\Big({\widetilde \phi}^{++}(\xi) - {\widetilde \phi}^{-+}(\xi)\Big) - 
\Big({\widetilde \phi}^{+-}(\xi) - {\widetilde \phi}^{--}(\xi)\Big)  =\frac{i}{2}
\widetilde{\big( \frac{\psi_{Bi}^2}{ \psi_{Ai}^2 }\big)}(\xi - \frac{8i}{3})
$$
for $\displaystyle \xi \in ]\frac{8i}{3}, \frac{12i}{3}[$. This reads
also:
\begin{equation}\label{2stcons}
\Big( {\widetilde \phi}^{++}(\xi) - {\widetilde \phi}^{+-}(\xi) \Big) -
\Big( {\widetilde \phi}^{-+}(\xi) - {\widetilde \phi}^{--}(\xi) \Big)
=\frac{i}{2} \widetilde{\big(\frac{\psi_{Bi}^2}{ \psi_{Ai}^2 }\big)}(\xi - \frac{8i}{3}), 
\hspace{5mm}
\xi \in ]\frac{8i}{3}, \frac{12i}{3}[.
\end{equation}
Comparing (\ref{1stcons}) and (\ref{2stcons}) we get that, for
$\displaystyle \xi \in ]\frac{8i}{3}, \frac{12i}{3}[$,
\begin{equation}\label{3stcons}
\left\{
\begin{array}{l}
\displaystyle {\widetilde \phi}^{++}(\xi) - {\widetilde \phi}^{+-}(\xi) =
\frac{i}{4} \widetilde{\big(\frac{\psi_{Bi}^2}{ \psi_{Ai}^2
  }\big)}(\xi - \frac{8i}{3}) \\
\\
\displaystyle {\widetilde \phi}^{-+}(\xi) - {\widetilde \phi}^{--}(\xi)
= - \Big( {\widetilde \phi}^{++}(\xi) - {\widetilde \phi}^{+-}(\xi)
\Big),
\end{array}
\right.
\end{equation}
while the comparison of (\ref{0stcons}) with (\ref{0stconsbis}) gives
\begin{equation}\label{4stcons}
\mbox{res}_{\frac{8i}{3}}  {\widetilde \phi}^{+}(\xi) = \frac{i}{4}, \hspace{5mm}
\mbox{res}_{\frac{8i}{3}}   {\widetilde \phi}^{-}(\xi) = -\frac{i}{4}.
\end{equation}

\begin{figure}[thp]
\begin{center}
\includegraphics[width=2.8in]{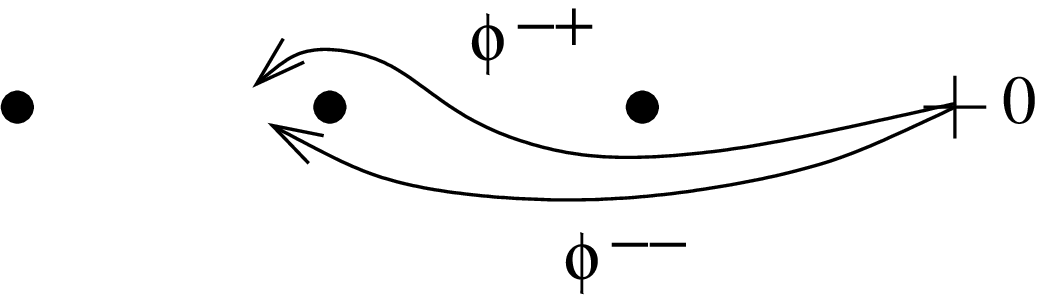} 
\caption{
\label{fig:Exemple14}}
\end{center}
\end{figure}

\subsubsection{Stokes automorphism}

\begin{figure}[thp]
\begin{center}
\includegraphics[width=3.0in]{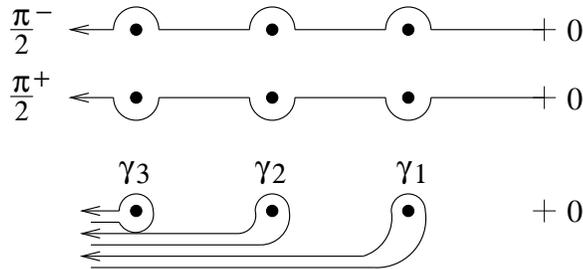} 
\caption{Comparing left and right Borel-resummation in the direction
  $\pi/2$ (we have rotated the picture). The bullets are the points
$4in/3$, $n \in \mathbb{N}^\star$.
\label{fig:figart5}}
\end{center}
\end{figure}

The above analysis  allows us to understand the Stokes phenomenon in the singular
direction $\pi/2$. We can write
\begin{equation}\label{leftrightgen}
\mbox{\sc  s}_{\frac{\pi}{2}^-}  \phi  (x)  = 
\mbox{\sc  s}_{\frac{\pi}{2}^+}  \phi  (x)  + \sum_{n \geq 1} 
\int_{\gamma_n} \widetilde{ \phi} (\xi)
e^{-x \xi} \, d\xi
\end{equation}
where the paths $\gamma_n$ are drawn on Fig. \ref{fig:figart5}. From what precedes,
$$\int_{\gamma_1} \widetilde{ \phi} (\xi)
e^{-x \xi} \, d\xi = e^{-\frac{4i}{3}x}\mbox{\sc  s}_{\frac{\pi}{2}^+} \frac{1}{2}\frac{\psi_{Bi}}{
  \psi_{Ai} }(x), 
\hspace{5mm}  
\int_{\gamma_2} \widetilde{ \phi} (\xi)
e^{-x \xi} \, d\xi = e^{-\frac{8i}{3}x} \mbox{\sc  s}_{\frac{\pi}{2}^+} \frac{i}{4} 
\frac{\psi_{Bi}^2}{ \psi_{Ai}^2 } (x).$$

To get the whole picture, there is no need to do things by hand as we
have previously done. We introduce a definition: 

\begin{Definition}
The alien derivation in the direction  $\pi/2$ is 
\begin{equation}\label{Deltadirect}
\underline{\Delta}^x_\frac{\pi}{2} = \sum_{n \in \mathbb{N}^\star} \dot  \Delta^x_{\frac{4i}{3}n}, 
\end{equation} 
where $\dot \Delta^x_\omega = e^{-\omega x}
\Delta^x_\omega$  is the pointed alien derivation at $\omega \in
\mathbb{C}^\star$. 
\end{Definition}
It can be shown \cite{Ec81-1, CNP2} that:
\begin{Theorem}
One has 
\begin{equation}\label{AutoStokes}
\mbox{\sc  s}_{\frac{\pi}{2}^-}  = \mbox{\sc  s}_{\frac{\pi}{2}^+}
\circ \mathfrak{S}_\frac{\pi}{2} \hspace{5mm} \mbox{where}   
\hspace{5mm} \mathfrak{S}_\frac{\pi}{2} = \exp (\underline{\Delta}^x_\frac{\pi}{2}) = 
\sum_{n=0}^\infty \frac{ (\underline{\Delta}^x_\frac{\pi}{2})^n }{n!}.
\end{equation}
The operator $\mathfrak{S}_\frac{\pi}{2}$ is the {\em Stokes
  automorphism in the direction  $\pi/2$}.
\end{Theorem}
In this theorem, by automorphism we mean automorphism of resurgent algebras
(we do not precise here  these algebras, see, e.g., \cite{CNP2}).\\
From (\ref{EQaphi}) we know that:
$${\underline \Delta}^x_{(\pi/2)} \phi =
 \frac{1}{2}\frac{\psi_{Bi}}{ \psi_{Ai} }
 e^{-\frac{4i}{3}x}.$$
Using (\ref{equpsiaibi}) one easily sees that for every  $k \geq 2$, 
$\Delta^x_{\frac{4ki}{3}} \Big( \Delta^x_{\frac{4i}{3}} \phi
\Big)=0$, so that $(\underline{\Delta}^x_{(\pi/2)})^2 \phi$ reduces to 
 $ e^{-\frac{8i}{3}x} \Delta^x_{\frac{4i}{3}} \Big( \Delta^x_{\frac{4i}{3}} \phi \Big)$:
$$(\underline{\Delta}^x_{(\pi/2)})^2 \phi = 
\frac{i}{2}\left( \frac{\psi_{Bi}}{ \psi_{Ai} }
 e^{-\frac{4i}{3}x} \right)^2.$$
More generally, for $n \geq 1$,
$$(\underline{\Delta}^x_{(\pi/2)})^{n} \phi= 
- \frac{i}{2} \Gamma (n) \left( i \frac{\psi_{Bi}}{ \psi_{Ai} }
 e^{-\frac{4i}{3}x} \right)^{n}.$$
Thus ${\mathfrak S}_{(\pi/2)} \, \phi$ is the following {\em
  transseries}: 
\begin{equation}\label{Act1Auto}
{\mathfrak S}_{(\pi/2)} \, \phi = 
\phi - \frac{i}{2} \sum_{n \geq 1} \frac{1}{n}\left(i \frac{\psi_{Bi}}{
    \psi_{Ai} } \right)^n e^{-\frac{4in}{3}x}
 = 
\phi + \frac{i}{2} \ln \left(1
-i \frac{\psi_{Bi}}{ \psi_{Ai} } e^{-\frac{4i}{3}x}\right).
\end{equation}
Of course one defines the {\em Stokes
  automorphism in the direction  $-\pi/2$ } (the other singular
direction)  in a similar way. The calculation gives
\begin{equation}\label{Act1Auto2}
{\mathfrak S}_{(-\pi/2)} \, \phi = 
\phi + \frac{i}{2} \sum_{n \geq 1} \frac{1}{n}\left( i \frac{\psi_{Ai}}{
    \psi_{Bi} } \right)^n e^{\frac{4in}{3}x}
 = 
\phi - \frac{i}{2} \ln \left(1
-i \frac{\psi_{Ai}}{ \psi_{Bi} } e^{\frac{4i}{3}x}\right).
\end{equation}

\subsection{Resurgent structure for the implicit function}\label{ResForX}

We now return to the resurgent formal series expansion $X(t)$
implicitly  defined
by equation (\ref{iterX2}). We would like to analyze its resurgent
structure from what we now  know concerning $\phi$. This requires the
following theorem, in complement with  theorem \ref{compose}:

\begin{Theorem}\label{composebis}
We assume that $\varphi_1(t)$ and  $\varphi_2(x)$ 
are resurgent formal series expansions. We assume furthermore that
$\varphi_1$ is small and we note $\varphi(t) = t + \varphi_1(t)$. Then for any
$\omega \in \mathbb{C}^\star$,
\begin{equation}\label{Erreur}
{\dot \Delta}^t_\omega  \Big( \varphi_2  \circ  \varphi \Big) (t) = 
({\dot \Delta}^x_\omega \varphi_2) \circ \varphi (t) +
\Big( {\dot \Delta}^t_\omega  \varphi (t) \Big) \Big( \varphi_2^\prime \circ
 \varphi (t) \Big).
\end{equation}
\end{Theorem}

In this theorem $^\prime$ means the usual derivation, while
$\displaystyle \Delta^t_\omega  \varphi (t) = \Delta^t_\omega
\varphi_1 (t)$.  
One can also translate (\ref{Erreur}) in term of usual
alien derivatives:
\begin{equation}\label{Erreurbis}
\Delta^t_\omega  \Big( \varphi_2  \circ  \varphi \Big) (t) = 
e^{-\omega \varphi_1(t)}\Big[(\Delta^x_\omega \varphi_2) \circ \varphi (t)\Big] +
\Big( \Delta^t_\omega  \varphi (t) \Big) \Big( \varphi_2^\prime \circ
 \varphi (t) \Big).
\end{equation}

Up to the knowledge of the author, theorem \ref{compose} only appears in
\cite{DP99} (with a misprint) with no explanation, so that it is perhaps worth to
give here at least a sketch of proof.

\begin{proof}
We recall that the composition function
$\varphi_2 \circ \varphi$ is  defined by 
$$\varphi_2 \circ \varphi (t)  = \sum_{n=0}^\infty \frac{\varphi_1^n(t)}{n!} 
 \frac{d^n \varphi_2 }{dx^n}(t) .$$
We recall also the following result \cite{Ec81-1, CNP2}: 
\begin{Theorem}\label{commutesalien}
The pointed alien derivation $\dot \Delta^t_\omega = e^{-\omega t}
\Delta^t_\omega$ commutes with the usual derivation $\displaystyle \frac{d}{dt}$.
\end{Theorem}
At a formal level this theorem and the Leibniz chain rule imply that, 
for $\omega \in \mathbb{C}^\star$,
$$\dot{\Delta}^t_\omega \Big( \varphi_2 \circ \varphi \Big)(t)  = 
\sum_{n=1}^\infty  \big( \dot{\Delta}^t_\omega \varphi_1 (t)\big)
\frac{\varphi_1^{n-1}(t)}{(n-1)!} 
  \frac{d^n \varphi_2 }{dx^n}(t) + 
\sum_{n=0}^\infty \frac{\varphi_1^n(t)}{n!}   \Big(\frac{d^n }{dx^n}\big(
\dot{\Delta}^x_\omega \varphi_2 \big)\Big) (t),$$
which is just
$$
{\dot \Delta}^t_\omega  \Big( \varphi_2  \circ  \varphi \Big) (t) = 
\Big( {\dot \Delta}^t_\omega  \varphi (t) \Big) \Big( \varphi_2^\prime \circ
 \varphi (t) \Big) + ({\dot \Delta}^x_\omega \varphi_2) \circ \varphi (t).
$$
To justify this formal reasoning, one just need to go back to  the proof
of the resurgence of the composition function $\varphi_2 \circ
\varphi$ (see \cite{CNP2}) and to translate the alien derivation in the Borel plane. 
\end{proof}

We apply theorem \ref{compose} to the implicit equation defining $X(t) = t + \X(t)$ (see
theorem \ref{theomp1}): for
$\omega \in \mathbb{C}^\star$, 
$$
\begin{array}{ll}
\displaystyle  \Delta^t_\omega  X  & \displaystyle  = \frac{3}{2}
\Delta^t_\omega \big( \phi \circ X \big) \\
 & \\
 & \displaystyle =   \frac{3}{2} \Big[  e^{-\omega \X(t)} ( \Delta^x_\omega \phi ) \circ X
 + \Delta^t_\omega  X ( \phi^\prime \circ X )\Big].
\end{array}
$$
Since $\phi^\prime$ is small, $2 - 3\phi^\prime \circ X$ is
invertible, its inverse being a resurgent formal series expansions. We
thus obtain:
$$\Delta^t_\omega  X = \frac{3}{2 - 3\phi^\prime \circ X} 
e^{-\omega \X(t)}  ( \Delta^x_\omega \phi ) \circ X.$$
Meanwhile (\ref{iterX2}) implies also 
$\displaystyle X^\prime = 1 + \frac{3}{2} X^\prime \big( \phi^\prime
\circ X \big)$ so that
$$X^\prime = \frac{2}{2 - 3\phi^\prime \circ X}$$
This finally gives:

\begin{Lemma}\label{beaulemma}
For $\omega \in \mathbb{C}^\star$,
$$\Delta^t_\omega  X = \frac{3}{2} X^\prime . e^{-\omega \X(t)} 
( \Delta^x_\omega \phi ) \circ X.$$
\end{Lemma}
With proposition \ref{REsforphi} this implies that:

\begin{Proposition}\label{REsforX}
The resurgent structure of $X(t)$ is governed by:
\begin{equation}\label{EQaX}
\left\{
\begin{array}{l}
\displaystyle  \Delta^t_{\frac{4i}{3}} X = \frac{3}{4} X^\prime
. e^{-(4i/3) \X(t)}  \left( \frac{\psi_{Bi}}{ \psi_{Ai} }\right) \circ X \\
\\
\displaystyle  \Delta^t_{-\frac{4i}{3}} X =-\frac{3}{4} X^\prime
. e^{(4i/3) \X(t)} \left( \frac{\psi_{Ai}}{ \psi_{Bi} }\right) \circ X \\
\\
\mbox{otherwise, } \, \Delta^t_{\omega} X =0
\end{array}
\right.
\end{equation}
\end{Proposition}
The calculation gives:
\begin{equation}\label{EQaX2}
\begin{array}{l}
\displaystyle \Delta^t_{\frac{4i}{3}} X  =  \frac{3}{4} -\frac{15}{128} t^{-2} +
  \frac{3765}{8192}t^{-4} - \frac{1360375}{262144} t^{-6} + \cdots
 =  \sum_{m=0}^\infty \frac{ c_{(m, \frac{4i}{3})} }{t^m}  \\
\\
\displaystyle \Delta^t_{-\frac{4i}{3}} X =  -\frac{3}{4} +\frac{15}{128} t^{-2} -
  \frac{3765}{8192}t^{-4} + \frac{1360375}{262144} t^{-6} - \cdots
 =  \sum_{m=0}^\infty \frac{ c_{(m, -\frac{4i}{3})} }{t^m} 
\end{array}
\end{equation}
where
$$ 
c_{(m, \frac{4i}{3})} = - c_{(m, -\frac{4i}{3})} \in \mathbb{R}.
$$
(This relies to the fact that $\psi_{Bi}
(x) = \psi_{Ai} (-x)$ and that $\psi_{Ai} (ix) \in
\mathbb{R}[[x^{-1}]]$).

Note that (\ref{EQaX}) implies that the alien derivative
$\displaystyle \underline{\Delta}^t_\frac{\pi}{2} X$ 
({\em resp.} $\displaystyle \underline{\Delta}^t_{-\frac{\pi}{2}} X$)
reduces to $\displaystyle e^{-\frac{4i}{3}t}  \Delta^t_{\frac{4i}{3}}X$  ({\em resp.} 
$\displaystyle e^{\frac{4i}{3}t}  \Delta^t_{-\frac{4i}{3}}X$).

To go further we remark that theorem \ref{commutesalien} translates into the fact that
\begin{equation}\label{commute}
\Delta^t_\omega \, \frac{d}{dt} = \big(-\omega + \frac{d}{dt} \big) \, \Delta^t_\omega .
\end{equation}
This being said, one uses again theorem \ref{composebis} to calculate 
$(\Delta^t_{\frac{4i}{3}})^2 X$: 
$$(\Delta^t_{\frac{4i}{3}})^2 X  = \frac{3}{4} \Delta^t_{\frac{4i}{3}} \left( 
X^\prime e^{-(4i/3) \X(t)} \left( \frac{\psi_{Bi}}{ \psi_{Ai}}\right) \circ X \right)$$
$$= \frac{3}{4} (\Delta^t_{\frac{4i}{3}} X^\prime)
e^{-(4i/3) \X(t)}  \left(\frac{\psi_{Bi}}{ \psi_{Ai}} \right) \circ X  
-i X^\prime (\Delta^t_{\frac{4i}{3}} X) e^{-(4i/3) \X(t)}\left(\frac{\psi_{Bi}}{ \psi_{Ai}} \right) \circ X
$$
$$+ 
\frac{3}{4}X^\prime e^{-(4i/3) \X(t)}\left(e^{-(4i/3) \X(t)}
\left(\Delta^x_{\frac{4i}{3}} \frac{\psi_{Bi}}{ \psi_{Ai}}\right) \circ X
+
\Delta^t_{\frac{4i}{3}} X .\left(\frac{\psi_{Bi}}{ \psi_{Ai}}
\right)^\prime \circ X
 \right).
$$
Since $\displaystyle \Delta^x_{\frac{4i}{3}} \frac{\psi_{Bi}}{
  \psi_{Ai}} = i \frac{\psi_{Bi}^2}{ \psi_{Ai}^2}$, and  
taking into account (\ref{EQaX}) and (\ref{commute}), one gets: 
$$(\Delta^t_{\frac{4i}{3}})^2 X  =$$
$$e^{-(8i/3) \X(t)}\Big[
\Big( \frac{9}{16}X^{\prime \prime} -  \frac{3}{2}i
\big(X^\prime\big)^2 + \frac{3}{4}i X^\prime \Big)
\left(\frac{\psi_{Bi}}{ \psi_{Ai}}\right)^2 \circ X 
+\frac{9}{8}\big(X^\prime\big)^2.\left[ \left(\frac{\psi_{Bi}}{
\psi_{Ai}}\right)^\prime   \frac{\psi_{Bi}}{ \psi_{Ai}}
\right] \circ X \Big].
$$
The calculation gives 
\begin{equation}\label{EQaX3}
(\Delta^t_{\frac{4i}{3}})^2 X =  -\frac{3}{4}i  +
\frac{15}{128}i \, t^{-2} +
  \frac{45}{256}\, t^{-3} - \frac{3765}{8192}i \, t^{-4} +
  \cdots   =    \sum_{m=0}^\infty \frac{ c_{(m, \frac{4i}{3}, \frac{4i}{3})} }{t^m}.
\end{equation}
In a similar way,
$$(\Delta^t_{-\frac{4i}{3}})^2 X  =$$
$$e^{(8i/3) \X(t)}\Big[
\Big( \frac{9}{16}X^{\prime \prime} +  \frac{3}{2}i
\big(X^\prime\big)^2 - \frac{3}{4}i X^\prime \Big)
\left(\frac{\psi_{Bi}}{ \psi_{Ai}}\right)^2 \circ X 
+\frac{9}{8}\big(X^\prime\big)^2.\left[ \left(\frac{\psi_{Bi}}{
\psi_{Ai}}\right)^\prime   \frac{\psi_{Bi}}{ \psi_{Ai}}
\right] \circ X \Big],
$$
\begin{equation}\label{EQaX3bis}
(\Delta^t_{-\frac{4i}{3}})^2 X =  
 \frac{3}{4}i  -
\frac{15}{128}i \, t^{-2} +
  \frac{45}{256}\, t^{-3} + \frac{3765}{8192}i \, t^{-4} +
  \cdots 
   =    \sum_{m=0}^\infty \frac{ c_{(m, -\frac{4i}{3}, -\frac{4i}{3})} }{t^m}.
\end{equation}
Here again 
$\displaystyle (\underline{\Delta}^x_\frac{\pi}{2})^2 X$ 
({\em resp.} $\displaystyle (\underline{\Delta}^x_{-\frac{\pi}{2}})^2 X$)
reduces to $\displaystyle e^{-\frac{8i}{3}t}  (
  \Delta^t_{\frac{4i}{3}})^2  X$  ({\em resp.} 
$\displaystyle e^{\frac{8i}{3}t}  (\Delta^t_{-\frac{4i}{3}})^2
X$). More generally we have
\begin{equation}\label{StoAutoX1}
\begin{array}{c}
\displaystyle {\mathfrak S}_{(\pi/2)} \, X =  X + \sum_{n \geq 1}
e^{-\frac{4i}{3}n t} \frac{1}{n !}  X_n(t)  \hspace{5mm}  
X_n = (\Delta^t_{\frac{4i}{3}})^n X\\
\\
\displaystyle  {\mathfrak S}_{(-\pi/2)} \, X =  X + \sum_{n \geq 1}
e^{\frac{4i}{3}n t} \frac{1}{n !}  X_{-n}(t)  \hspace{5mm}
X_{-n} = (\Delta^t_{-\frac{4i}{3}})^n X 
\end{array}
\end{equation}
For latter purpose (\S \ref{Hyper2}) it is also necessary to
calculate the alien derivatives $\Delta^t_{-\frac{4i}{3}} \big(
\Delta^t_{\frac{4i}{3}} X \big)$ and $\Delta^t_{+\frac{4i}{3}} \big(
\Delta^t_{-\frac{4i}{3}} X \big)$. The reasoning is quite analogous to
what we have done previously. One obtains:
$$\Delta^t_{-\frac{4i}{3}} \big( \Delta^t_{\frac{4i}{3}} X \big)   =
-\frac{3}{4}i \,X^\prime
- \frac{9}{16}X^{\prime \prime} 
$$
\begin{equation}\label{EQaX4}
\Delta^t_{-\frac{4i}{3}} \big( \Delta^t_{\frac{4i}{3}} X \big)  =
 -\frac{3}{4}i  +
\frac{15}{128}i \, t^{-2} -
  \frac{45}{256}\, t^{-3} - \frac{3765}{8192}i \, t^{-4} +
  \cdots 
  =   \sum_{m=0}^\infty \frac{ c_{(m, \frac{4i}{3}, -\frac{4i}{3})} }{t^m}.
\end{equation}
while
$$\Delta^t_{\frac{4i}{3}} \big( \Delta^t_{-\frac{4i}{3}} X \big)   =
\frac{3}{4}i \, X^\prime
- \frac{9}{16}X^{\prime \prime}  ,
$$
\begin{equation}\label{EQaX4bis}
\Delta^t_{\frac{4i}{3}} \big( \Delta^t_{-\frac{4i}{3}} X \big)  =
 \frac{3}{4}i  -
\frac{15}{128}i \, t^{-2} -
  \frac{45}{256}\, t^{-3} + \frac{3765}{8192}i \, t^{-4} +
  \cdots 
  =   \sum_{m=0}^\infty \frac{ c_{(m, -\frac{4i}{3}, \frac{4i}{3})} }{t^m}.
\end{equation}
Note that
\begin{equation}\label{StoAutoX11}
\begin{array}{c}
\displaystyle {\mathfrak S}_{(\pi/2)} \, X_1 =  X_1 + \sum_{n \geq 1}
e^{-\frac{4i}{3}n t} \frac{1}{n !} X_{(1,n)}(t)  \hspace{3mm}  \mbox{with} \hspace{3mm} 
X_{(1,1)} = \Delta^t_{\frac{4i}{3}} X_1 = (\Delta^t_{\frac{4i}{3}})^2 X\\
\\
\displaystyle  {\mathfrak S}_{(-\pi/2)} \, X_1 =  X_1 + \sum_{n \geq 1}
e^{\frac{4i}{3}n t} \frac{1}{n !} X_{(1,-n)}(t)  \hspace{3mm}  \mbox{with}
\hspace{3mm}  
X_{(1,-1)} = \Delta^t_{-\frac{4i}{3}} X_1 = \Delta^t_{-\frac{4i}{3}} \big( \Delta^t_{\frac{4i}{3}} X \big)
\end{array}
\end{equation}
while
\begin{equation}\label{StoAutoX12}
\begin{array}{c}
\displaystyle {\mathfrak S}_{(\pi/2)} \, X_{-1} =  X_{-1} + \sum_{n \geq 1}
e^{-\frac{4i}{3}n t} \frac{1}{n !} X_{(-1,n)}(t)  \hspace{3mm}  \mbox{with} \hspace{3mm} 
X_{(-1,1)} = \Delta^t_{\frac{4i}{3}} X_{-1} = \Delta^t_{\frac{4i}{3}} \big( \Delta^t_{-\frac{4i}{3}} X \big)\\
\\
\displaystyle  {\mathfrak S}_{(-\pi/2)} \, X_{-1} =  X_{-1} + \sum_{n \geq 1}
e^{\frac{4i}{3}n t} \frac{1}{n !} X_{(-1,-n)}(t)  \hspace{3mm}  \mbox{with} 
\hspace{3mm}  
X_{(-1,-1)} = \Delta^t_{-\frac{4i}{3}} X_{-1} = (\Delta^t_{-\frac{4i}{3}})^2 X
\end{array}
\end{equation}

\section{The zeros of the Airy function: hyperasymptotic  method}\label{section5}

We now show how the informations we have got about $X$ can be used in
hyperasymptotic calculations. Since this is the first paper where the 
relationships between resurgent theory and hyperasymptotics is done,
we shall detail  the constructions. However we refer to \cite{Old98}
for questions related to optimal truncations at each hyperasymptotic
level, and for remainder estimates.

\subsection{Level-0}\label{Hyper0} 
We start at the $0$ level, which  is just the
summation to the least term. From theorem \ref{theomp1} and referring for instance to
\cite{DelabRaso06-2}, one sees that for $ r \in ]0, 4/3[$, there exist
$A >0$, $B>0$ such that, for $\Re(t)> B $ et $ n\geq 0$,
\begin{equation}\label{pluspetit}
\begin{array}{c}
 \displaystyle
  \Big|\mbox{\sc  s}_0 X(t) - t - \sum_{k=0}^{n}
  \frac{c_k}{t^k}\Big| \leq R_{as}(r,A,B,n,t)\\
\\
 \displaystyle R_{as}(r, A,B,n,t) = Ae^{Br}  \frac{ n!  }{r^n} \frac{1}{|t|^{n}(\Re
    (t)-B)}.
\end{array}
\end{equation}
To minimize $R_{as}(r, A,B,n,xt$, one is brought to choose  $n =
\big[r|z|\big]$ as optimal truncation ($\big[ . \big]$ is the integer
part). For such a $n$ one evaluates in practice $ R_{as}(r, A,B,n,t)$
as 
$$ R_{as}(r, A,B,n,t) \sim \frac{|c_{n+1}|}{|t|^{n} \Re (t)}$$
for $n$ and $|x|$ large enough. However, since $c_n =0$ for $n$ even,
we shall use these estimates only for $n+1$ odd. The calculations made
with $\displaystyle n = \sup_{0<r<4/3} \big[r|t|\big]$ 
provide tables  \ref{table3} and \ref{table4}.

\begin{table}[thp]
\begin{center}
\begin{tabular}{|c|l|c|c|}
\hline
Optimal  $n$ & Estimates for $k_1$ & Error estimates & Real error\\
\hline
5 & -2.33863 & $0.147 \times 10^{-2}$ & $0.52\times 10^{-3}$\\
\hline
\end{tabular}
\caption{Calculation for the first zero $k_1$ of the Airy function by
  level-0 hyperasymptotics.  
\label{table3}}
\end{center}
\end{table}

\begin{table}[thp]
\begin{center}
\begin{tabular}{|c|l|c|c|}
\hline
Optimal  $n$ & Estimates for $k_3$ & Error estimates & Real error\\
\hline
17 & -5.52055982870 & $0.133 \times 10^{-8}$ & $0.60\times 10^{-9}$\\
\hline
\end{tabular}
\caption{Calculation for the first zero $k_3$ of the Airy function by
  level-0 hyperasymptotics.  
\label{table4}}
\end{center}
\end{table}

\subsection{Level-1}\label{Hyper1} 
For $\Re (t) > B$ ($B$ large enough), the Borel sum of $\displaystyle X(t) = 
t + \sum_{n=0}^\infty \frac{c_n}{t^n}$ reads 
$$\mbox{\sc  s}_0 X(t) = t + \int_0^{+\infty} e ^{-t\tau}
\widetilde{X}(\tau) \, d\tau,$$
where $\widetilde{X}(\tau)$ is the minor of $X(t)$ (since $c_0 = 0$,
see (\ref{expXt}). \\
For practical calculation  one can introduce a cut-off $b>0$ large enough in the
integral so that,  instead of working with the Borel-sum
$\mbox{\sc  s}_0 X(t)$ one considers the function
\begin{equation}\label{Sumb}
\mbox{\sc  s}_0^b X(t) = t + \int_0^{b} e ^{-t\tau}
\widetilde{X}(\tau) \, d\tau.
\end{equation}
This can be justified as follows. Since $X$ is Borel-resummable (in
the direction $0$) there exist $A>0$ and $B>0$ such that for $\tau \in
B_r$ (see Notation  \ref{Notation2}), $r>0$ small enough (here  $0< r <
4/3$), $|\widetilde{X}(\tau)| \leq A e^{B|\tau|}$. \\
For  $0 < \delta  < \pi/2$ and $\mu >1$ we note 
$$P_{\delta, \mu} (B) = \{t \in  \mathbb{C} \, / \,
|\arg(t)| \leq   \frac{\pi}{2} -\delta, \, |z| \geq 
\frac{\mu B}{\sin(\delta)}\}.$$
For $t \in P_{\delta, \mu} (B)$ we have
$\Re(t) - B \geq  (1-\frac{1}{\mu})\sin(\delta)|t| \geq (\mu-1)B$, therefore
$$\left| \int_b^{+\infty}\widetilde{X}(\tau)e^{-t \tau} \, d\tau
\right| \leq \frac{e^{(B - \Re(t))b} }{\Re (t) -B} \leq 
\frac{e^{-(1-\frac{1}{\mu}) b \sin(\delta)|t|} }{(\mu-1)B}.
$$
This ensures that, for $t \in P_{\delta, \mu} (B)$ and for $b >0$
large enough, $|\mbox{\sc  s}_0 X(t) - \mbox{\sc  s}_0^b X(t)|$ is
numerically negligeable. 

In what follows, it is worth to work with a family of such $\mbox{\sc
  s}_0^b X(t)$, $b$ large enough. For $t  \in P_{\delta, \mu}
(B)$, we shall note $\displaystyle \mbox{\sc  s}_0^{[]} X(t)$,
\begin{equation}\label{Sumbb}
\mbox{\sc  s}_0^{[]} X(t) = t + \int_0^{[0]} e ^{-t\tau}
\widetilde{X}(\tau) \, d\tau, \hspace{5mm} [0] = b e^{i 0}, \, b>0
\mbox{ large enough.}
\end{equation}
a member of this family. One defines $\mbox{\sc  s}_\theta^{[]}$ ({\em
  resp.} 
$\mbox{\sc  s}_{\theta +}^{[]}$, $\mbox{\sc  s}_{\theta -}^{[]}$) in
a similar way, calling these operators {\em pre-Borel-resummation} ({\em
  resp. right, left   pre-Borel-resummation}) in the direction
$\theta$. One of the main advantage of working with
pre-Borel-resummation is that a pre-Borel-sum extends 
analytically as an entire function.

\begin{figure}[thp]
\begin{center}
\includegraphics[width=3.2in]{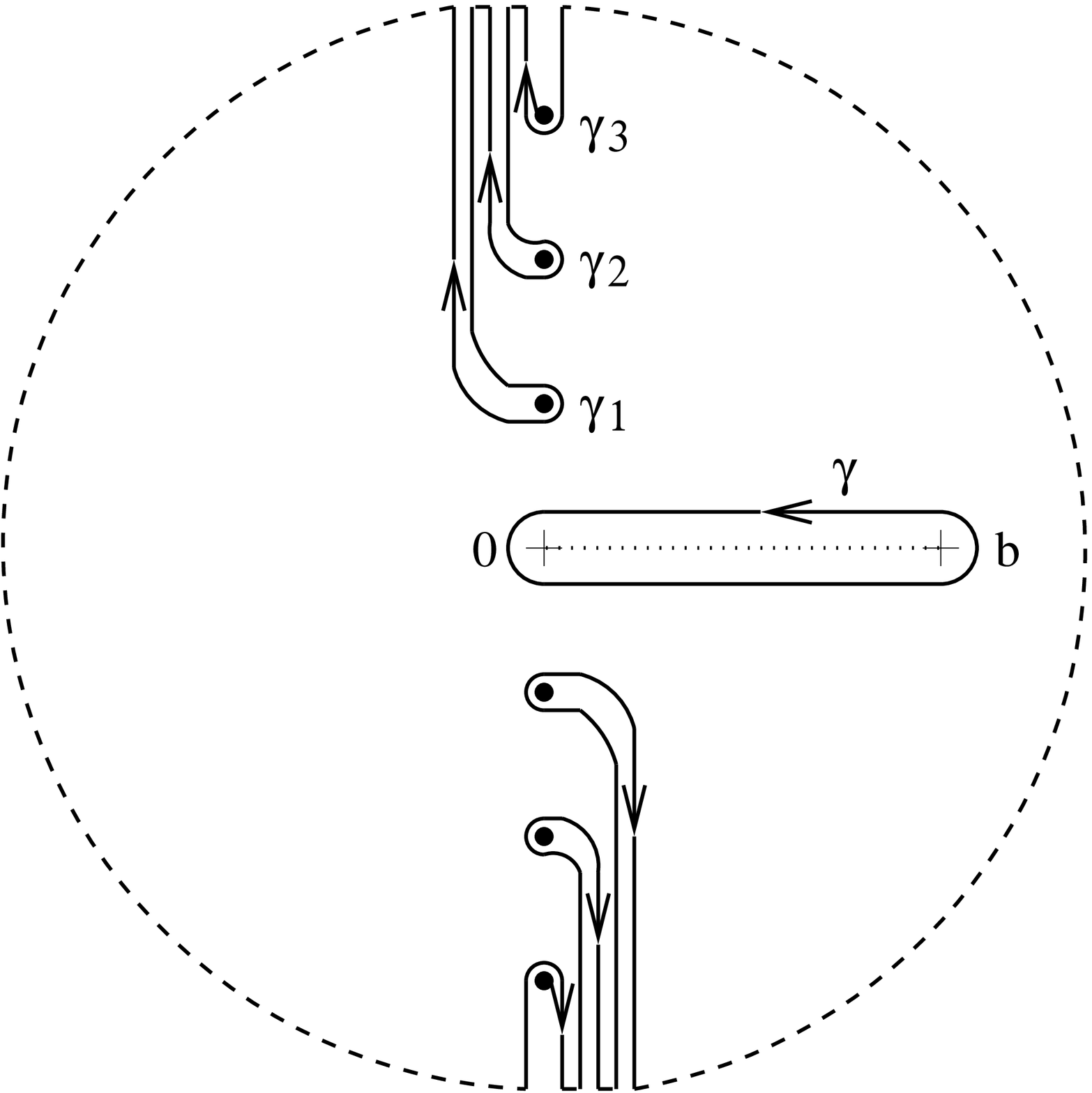} 
\caption{The contour $\gamma$ and its homotopic deformation. The
  bullets are the singular points $\omega_k = 4ik/3$ and 
$\omega_k^\star = - 4ik/3$, $k \in   \mathbb{N}^\star$. 
\label{fig:figart6}}
\end{center}
\end{figure}

In (\ref{Sumbb}) we represent  the function $\widetilde{X}(\tau)$ in
term of a Cauchy integral representation
\begin{equation}
\widetilde{X}(\tau) = \frac{1}{2i\pi} \oint  du
\frac{\widetilde{X} (u)}{u -\tau}.
\end{equation} 
By a  binomial expansion to the order of truncation $N_0$  one gets
the well-known Hermite formula for $\widetilde{X}(\tau)$ which, used
in (\ref{Sumb}), gives
\begin{equation}
\mbox{\sc  s}_0^{[]} X(t) = t+  \sum_{n=0}^{N_0-1} \frac{c_n}{t^n} + R(t, N_0)
\end{equation}
where
\begin{equation}
R(t, N_0) =\frac{1}{2i\pi t^{N_0}}
\int_0^{[\arg (t)]} dw e^{-w} w^{{N_0}-1}\int_{\gamma} du \frac{\widetilde{X} (u)}{(1- w/(tu))u^{N_0}}.
\end{equation}
The contour $\gamma$ encircles the line segment $[0,b]$ as in Fig \ref{fig:figart6}.
We then deform $\gamma$ as shown in Fig. \ref{fig:figart6}. By the
Cauchy theorem, on can write
\begin{equation}\label{hypeint1}
\begin{array}{ll}
\displaystyle R(t, {N_0}) = &  \displaystyle \frac{1}{2i\pi t^{{N_0}}} \sum_{k=1}^l 
\int_0^{[\arg (t)]} dw e^{-w} w^{{N_0}-1}\int_{\gamma_k} du \frac{\widetilde{X}
  (u)}{(1- w/(tu))u^{N_0}} \\
\\
 &\displaystyle  + 
\frac{1}{2i\pi t^{{N_0}}} \sum_{k=1}^l 
\int_0^{[\arg (t)]} dw e^{-w} w^{{N_0}-1}\int_{\gamma_k^\star} du \frac{\widetilde{X}
  (u)}{(1- w/(tu))u^{N_0}} \\
\\
& \displaystyle  + \frac{1}{2i\pi t^{{N_0}}}
\int_0^{[\arg (t)]} dw e^{-w} w^{{N_0}-1}\int_{C} du \frac{\widetilde{X}
  (u)}{(1- w/(tu))u^{N_0}}.
\end{array}
\end{equation}
In (\ref{hypeint1}) the $\gamma_k$ are the bounded paths drawn on Fig.
\ref{fig:figart6} and the $\gamma_k^\star$ are their complex
conjugates. The number of terms $l$ in each sum depends on the chosen
cut-off. 
The path $C$ consists in the remaining  arcs (in dotted
lines on Fig. \ref{fig:figart6}). As shown in \cite{Old98}, the path $C$ gives 
a  contribution to $R(t,{N_0})$ which  can be bounded away to an exponential level smaller 
than the one to which the hyperasymptotics is eventually taken, i.e., 
less than $\exp(-M|t|)$ for any chosen $M>0$, so that we shall forget
this term in what follows.

For each $\gamma_i$ ({\em resp.}  $\gamma_i^\star$) we make the change
of variable   $w = vu / \omega_k$ ({\em resp.} $w = vu /
\omega_k^\star$) where $\omega_k = 4ik/3$. Equality (\ref{hypeint1})
becomes (forgetting the contribution of the path $C$ as we explained):
\begin{equation}\label{hypeint2}
\begin{array}{ll}
\displaystyle R(t, {N_0}) = &  \displaystyle \frac{1}{2i\pi t^{{N_0}}} \sum_{k=1}^{l}  \frac{1}{\omega_k^{{N_0}}}
\int_0^{[\arg (t)]} dv \,  \int_{\gamma_k} du \, e^{-vu/\omega_k} v^{{N_0}-1}\frac{v^{{N_0}-1}}{1 - v/(t \omega_k)}   \widetilde{X}(u).
\\
 &\displaystyle  + 
\displaystyle \frac{1}{2i\pi t^{{N_0}}} \sum_{k=1}^{l}  \frac{1}{{\omega_k^\star}^{{N_0}}}
\int_0^{[\arg (t)]} dv \,  
\int_{\gamma_k^\star} du \, e^{-vu/\omega_k} \frac{v^{{N_0}-1}}{1 - v/(t \omega_k^\star)}   \widetilde{X}(u).
\end{array}
\end{equation}
Up to their orientations (and the cut-off), the contours $\gamma_k$
({\em resp.} $\gamma_k^\star$) are
those used when one compares left and right Borel-resummation in the
direction $\pi/2$ (see
Fig. \ref{fig:figart5}) ({\em resp.} $-\pi/2$). Using the known action of the Stokes
automorphisms $\displaystyle {\mathfrak S}_{(\pi/2)}$ and $\displaystyle {\mathfrak S}_{(-\pi/2)}$
on $X$ (see (\ref{StoAutoX1})) and making the convenient change of variable 
$\tau = u - \omega_k$ ({\em resp.} $\tau = u - \omega_k^\star$) in
each term, one obtains
\begin{equation}\label{hypeint3}
\begin{array}{ll}
\displaystyle R(t, {N_0}) = &  
\displaystyle -\frac{1}{2i\pi t^{{N_0}}} \sum_{k=1}^{l}  \frac{1}{\omega_k^{{N_0}}}
\int_0^{[\arg (t)]} dv \, e^{-v} \frac{v^{{N_0}-1}}{1 - v/(t \omega_k)} 
\mbox{\sc  s}_{(\pi/2)+}^{[]} \frac{1}{k !} X_k(v/\omega_k)
\\ 
 &\displaystyle  - 
\displaystyle \frac{1}{2i\pi t^{{N_0}}} \sum_{k=1}^{l}  \frac{1}{{\omega_k^\star}^{{N_0}}}
\int_0^{[\arg (t)]} dv \, e^{-v} \frac{v^{{N_0}-1}}{1 - v/(t
  \omega_k^\star)}   \mbox{\sc  s}_{(-\pi/2)+}^{[]} \frac{1}{k !} X_{-k}(v/\omega_k^\star).
\end{array}
\end{equation}
Note that, since one works with pre-Borel-resummation, $\mbox{\sc
  s}_{(\pi/2)+}^{[]} X_k$ and $\mbox{\sc  s}_{(-\pi/2)+}^{[]}
X_{-k}$ extend as entire functions, so that (\ref{hypeint3})
is well-defined.\\
Equation (\ref{hypeint3}) may be written in the following way by 
suitable changes of variables $\tau = v/\omega_k$ and $\tau = v/\omega_k^\star$.
\begin{equation}\label{hypeint3bis}
\begin{array}{ll}
\displaystyle R(t, {N_0}) = &  
\displaystyle -\frac{1}{2i\pi t^{{N_0-1}}} \sum_{k=1}^{l}  
\int_0^{[\arg (t) - \arg (\omega_k)]} \hspace{-5mm}  d\tau \, e^{-\omega_k \tau} 
\frac{\tau^{{N_0}-1}}{t -\tau} 
\mbox{\sc  s}_{(\pi/2)+}^{[]}\frac{1}{k !} X_k(\tau)
\\ 
 &\displaystyle  - 
\displaystyle \frac{1}{2i\pi t^{{N_0-1}}} \sum_{k=1}^{l}  
\int_0^{[\arg (t)-\arg (\omega_k^\star)]} \hspace{-5mm} d\tau \, e^{-\omega_k^\star \tau} 
\frac{\tau^{{N_0}-1}}{t -\tau} 
\mbox{\sc  s}_{(-\pi/2)+}^{[]} \frac{1}{k !} X_{-k}(\tau).
\end{array}
\end{equation}

The resurgence formula (\ref{hypeint3}) (in the sense of
Dingle-Berry-Howls \cite{BeH91, Din73}) is the key-point in hyperasymptotic theory.

The algorithm for the level-1 hyperasymptotics is now as follows (see \cite{Old98}):
\begin{itemize}
\item Only the seen (adjacent) singularities $\omega_1 = 4i/3$ and
  $\omega_1^\star = -4i/3$ play a role. This means that we write
\begin{equation}\label{hypeint4}
\begin{array}{ll}
\displaystyle R(t, {N_0}) = &  
\displaystyle -\frac{1}{2i\pi t^{{N_0-1}}}  
\int_0^{[\arg (t) - \arg (\omega_1)]} \hspace{-5mm} d\tau \, e^{-\omega_1 \tau} 
\frac{\tau^{{N_0}-1}}{t -\tau} 
\mbox{\sc  s}_{(\pi/2)+}^{[]} X_1(\tau)
\\ 
 &\displaystyle  - 
\displaystyle \frac{1}{2i\pi t^{{N_0-1}}} 
\int_0^{[\arg (t)-\arg (\omega_1^\star)]} \hspace{-5mm} d\tau \, e^{-\omega_1^\star \tau} 
\frac{\tau^{{N_0}-1}}{t -\tau} 
\mbox{\sc  s}_{(-\pi/2)+}^{[]} X_{-1}(\tau).
\end{array}
\end{equation}
modulo a remainder which will be negligeable at this level 1.
\item We replace each right Borel-sum  $\mbox{\sc
  s}_{(\pi/2)+}^{[]} X_1 (\tau)$ and $\mbox{\sc  s}_{(-\pi/2)+}^{[]}
\overline{X}_1 (\tau)$ by their truncated asymptotic
expansions. For reasons of symmetries, the order of truncation $N_1
\leq N_0$ will be
the same for each of these expansions.  One then extends  the bounded
contours of integration up to infinity. Putting all pieces together,
using (\ref{StoAutoX1}) and (\ref{EQaX2}), one thus obtains
\begin{equation}\label{hypeint5}
\begin{array}{ll}
\displaystyle \mbox{\sc  s}_0 X(t) = &  \displaystyle  t+  \sum_{n=0}^{N_0-1}
\frac{c_n}{t^n} \\
&   \displaystyle -\frac{1}{2i\pi t^{N_0-1}} 
 \sum_{n=0}^{N_1-1} c_{(n, \frac{4i}{3})}
\int_0^{ \infty e^{-i\arg(\omega_1)}} \hspace{-5mm} d\tau \, e^{-\omega_1
  \tau}  \frac{\tau^{N_0-n-1}}{t -\tau}
\\ 
 &\displaystyle  - \frac{1}{2i\pi t^{N_0-1}} 
 \sum_{n=0}^{N_1-1} c_{(n, -\frac{4i}{3})}
\int_0^{ \infty e^{-i\arg(\omega_1^\star)}} \hspace{-5mm} d\tau \, e^{-\omega_1^\star
  \tau}  \frac{\tau^{N_0-n-1}}{t -\tau} + R(N_0, N_1, t)
\end{array}
\end{equation} 
\item Taking into account that the seen singularities from 
$\omega_1$ ({\em resp.} $\omega_1^\star$) are at a distance
$|\omega_1|$, and  for reasons explained in \cite{Old98}, the choices $N_0 =
2|\omega_1| |t[ = 8/3|t|$ and $N_1 = |\omega_1| |t[ = 4/3|t|$  give
optimal truncations, for which the remainder term behaves like 
$R(N_0, N_1, t) = \exp\big( -N_0 |t| \big) O(1)$, for 
$t  \in P_{\delta, \mu} (B)$.
\end{itemize}
Note that one can write (\ref{hypeint5}) using the  
the canonical  hyperterminants \cite{BeH91, Ho92, Ho97, Old98},
defined by: 
\begin{equation}\label{Hypert}
\left\{
\begin{array}{l}
F^{(0)}(t)=1\\
\\
\displaystyle F^{(1)}\left(t; \begin{array}{c}
M_0\\ \sigma_0
\end{array}
\right)= \int_0^{\infty e^{-i\theta_0}} \hspace{-5mm} d\tau_0  \, e^{-\sigma_0 \tau_0}
\frac{\tau_0^{M_0-1}}{t-\tau_0}, \hspace{5mm} \theta_0= \arg (\sigma_0) \\
\\
F^{(l+1)}\left(t; \begin{array}{ccc}
M_0 , &\cdots , & M_l \\
\sigma_0 , &\cdots , & \sigma_l
\end{array}
\right)= \\
\hspace{20 mm}\displaystyle  \int_0^{\infty e^{-i\theta_0}}
\hspace{-5mm} d\tau_0 \cdots 
\int_0^{\infty e^{-i\theta_l}} \hspace{-5mm} d\tau_l  \,
e^{-(\sigma_0 \tau_0  + \cdots + \sigma_l \tau_l )}
\frac{\tau_0^{M_0-1} \cdots \tau_l^{M_l-1}}{(t-\tau_0)(\tau_0-\tau_1)
  \cdots (\tau_{l-1} -\tau_l)}\\
\hspace{0 mm}\displaystyle  (\theta_i= \arg (\sigma_i)).
\end{array}
\right.
\end{equation}

\begin{rem}
With formula (\ref{EQaX2}) we have noted that 
$$ 
c_{(m, \frac{4i}{3})} = - c_{(m, -\frac{4i}{3})} \in \mathbb{R}.
$$
Since  formula (\ref{hypeint5}) reads also as
\begin{equation}\label{hypeint6}
\begin{array}{ll}
\displaystyle \mbox{\sc  s}_0 X(t) = &  \displaystyle  t+  \sum_{n=0}^{N_0-1}
\frac{c_n}{t^n} \\
&   \displaystyle +\frac{1}{2 \pi t^{N_0-1}} 
 \sum_{n=0}^{N_1-1} (-i)^{N_0-n-1} c_{(n, \frac{4i}{3})}  
\int_0^{+\infty} dx \, e^{-4x/3}   \frac{x^{N_0-n-1}}{t +ix}
\\ 
 &\displaystyle  - \frac{1}{2\pi t^{N_0-1}} 
 \sum_{n=0}^{N_1-1} i^{N_0-n-1} c_{(n, -\frac{4i}{3})}
\int_0^{+\infty} dx \, e^{-4x/3}   \frac{x^{N_0-n-1}}{t -ix}  + R(N_0,
N_1, t), 
\end{array}
\end{equation} 
we see that the realness of $\displaystyle \mbox{\sc  s}_0 X(t)$ is
well preserved by the level-1 hyperasymptotics.
\end{rem}

We turn now to numerical experiments. Formula (\ref{hypeint5}) 
 give tables \ref{table5} and \ref{table6}.

\begin{table}[thp]
\begin{center}
\begin{tabular}{|c|l|c|}
\hline
Optimal  $N$ & Estimates for $k_1$ &  Real error\\
\hline
$N_0 = 9$, $N_1 =5$  & -2.33810834 & $0.93 \times 10^{-6}$ \\
\hline
\end{tabular}
\caption{Calculation for the first zero $k_1$ of the Airy function by
  level-1 hyperasymptotics.  
\label{table5}}
\end{center}
\end{table}

\begin{table}[thp]
\begin{center}
\begin{tabular}{|c|l|c|}
\hline
Optimal  $N$ & Estimates for $k_3$ &  Real error\\
\hline
$N_0 = 35$, $N_1 =17$  & -5.520559828095551059172 & $0.42 \times 10^{-19}$ \\
\hline
\end{tabular}
\caption{Calculation for the first zero $k_3$ of the Airy function by
  level-1 hyperasymptotics.  
\label{table6}}
\end{center}
\end{table}

\subsection{Level-2}\label{Hyper2}

What we have done at the level-1 can be repeated. We just detail here
how the informations got from the resurgence analysis can be used in
this context, referring to \cite{Old98} for what concerns the
questions of optimal truncations and remainder estimates.

We go back to (\ref{hypeint3bis}) with $l=2$. For $k=2$, one just
replace the pre-Borel-sums $ \displaystyle \mbox{\sc  s}_{(\pi/2)+}^{[]} \frac{1}{2 !} X_2(\tau)$
and $ \displaystyle \mbox{\sc  s}_{(-\pi/2)+}^{[]} \frac{1}{2 !} X_{-2}(\tau)$ by their
truncated asymptotic expansions. With the notations (\ref{StoAutoX1}),
(\ref{EQaX3}), (\ref{EQaX3bis}), (\ref{Hypert}), this gives:

\begin{equation}\label{hypeint3ter}
\begin{array}{l}
\displaystyle \mbox{\sc  s}_0 X(t) =  t+
\sum_{n=0}^{N_0-1} \frac{c_n}{t^n} \\
 \displaystyle -\frac{1}{2i\pi t^{N_0-1}} \sum_{n=0}^{N_2-1} \frac{
  c_{(n, \frac{4i}{3}, \frac{4i}{3})}}{2} 
F^{(1)}\left(t; \begin{array}{c}
N_0-n\\ \omega_2
\end{array} \right) +
\frac{c_{(n, -\frac{4i}{3}, -\frac{4i}{3})}}{2}
F^{(1)}\left(t; \begin{array}{c}
N_0-n\\ \omega_2^\star
\end{array} \right)
\\ 
\displaystyle -\frac{1}{2i\pi t^{{N_0-1}}}  
\int_0^{[\arg (t) - \arg (\omega_1)]} \hspace{-5mm} d\tau \, e^{-\omega_1 \tau} 
\frac{\tau^{{N_0}-1}}{t -\tau} 
\mbox{\sc  s}_{(\pi/2)+}^{[]} X_1(\tau)\\ 
  - 
\displaystyle \frac{1}{2i\pi t^{{N_0-1}}} \sum_{k=1}^{l}  
\int_0^{[\arg (t)-\arg (\omega_1^\star)]} \hspace{-5mm} d\tau \, e^{-\omega_1^\star \tau} 
\frac{\tau^{{N_0}-1}}{t -\tau} 
\mbox{\sc  s}_{(-\pi/2)+}^{[]} X_{-1}(\tau) + R(N_0, N_2, t).
\end{array}
\end{equation}

For $\mbox{\sc  s}_{(\pi/2)+}^{[]} X_1(\tau)$ and $\mbox{\sc
  s}_{(-\pi/2)+}^{[]} X_{-1}(\tau)$ we copy what we have done at the
level-1 hyperasymptotics: using (\ref{StoAutoX11}) and
(\ref{StoAutoX12}) we get:
\begin{equation}\label{hypeint7}
\begin{array}{l}
\displaystyle \mbox{\sc  s}_{(\pi/2)+}^{[]} X_1(\tau)
 =  \sum_{n=0}^{N_1-1} \frac{c_{(n,4i/3)}}{\tau^n} \\
  \displaystyle -\frac{1}{2i\pi \tau^{N_1-1}} 
 \sum_{n=0}^{N_2-1} c_{(n, \frac{4i}{3}, \frac{4i}{3})}
\int_0^{ \infty e^{-i\arg(\omega_1)}} \hspace{-5mm} d\tau_1 \, e^{-\omega_1
  \tau_1}  \frac{\tau_1^{N_1-n-1}}{\tau -\tau_1}
\\ 
\displaystyle  - \frac{1}{2i\pi \tau^{N_1-1}} 
 \sum_{n=0}^{N_2-1} c_{(n,\frac{4i}{3}, -\frac{4i}{3})}
\int_0^{ \infty e^{-i\arg(\omega_1^\star)}} \hspace{-5mm} d\tau_1 \, e^{-\omega_1^\star
  \tau_1}  \frac{\tau_1^{N_1-n-1}}{\tau -\tau_1} + R(N_1, N_2, \tau)
\end{array}
\end{equation} 
and
\begin{equation}\label{hypeint8}
\begin{array}{l}
\displaystyle \mbox{\sc  s}_{(-\pi/2)+}^{[]} X_{-1}(\tau)
 = \sum_{n=0}^{N_1-1}
\frac{c_{(n,-4i/3)}}{\tau^n} \\
  \displaystyle -\frac{1}{2i\pi \tau^{N_1-1}} 
 \sum_{n=0}^{N_2-1} c_{(n, -\frac{4i}{3}, \frac{4i}{3})}
\int_0^{ \infty e^{-i\arg(\omega_1)}} \hspace{-5mm} d\tau_1 \, e^{-\omega_1
  \tau_1}  \frac{\tau_1^{N_1-n-1}}{\tau -\tau_1}
\\ 
\displaystyle  - \frac{1}{2i\pi \tau^{N_1-1}} 
 \sum_{n=0}^{N_2-1} c_{(n,-\frac{4i}{3}, -\frac{4i}{3})}
\int_0^{ \infty e^{-i\arg(\omega_1^\star)}} \hspace{-5mm} d\tau_1 \, e^{-\omega_1^\star
  \tau_1}  \frac{\tau_1^{N_1-n-1}}{\tau -\tau_1} + R(N_1, N_2, \tau)
\end{array}
\end{equation} 
Plugging (\ref{hypeint7}) and (\ref{hypeint8}) in (\ref{hypeint3ter})
we obtain:

\begin{equation}\label{hypeint10}
\begin{array}{l}
\displaystyle \mbox{\sc  s}_0 X(t) =  t+
\sum_{n=0}^{N_0-1} \frac{c_n}{t^n} \\
 \displaystyle -\frac{1}{2i\pi t^{N_0-1}} \sum_{n=0}^{N_1-1}  c_{(n, \frac{4i}{3})}
F^{(1)}\left(t; \begin{array}{c}
N_0-n\\ \omega_1
\end{array} \right) + c_{(n, -\frac{4i}{3})}
F^{(1)}\left(t; \begin{array}{c}
N_0-n\\ \omega_1^\star
\end{array} \right)\\
 \displaystyle -\frac{1}{2i\pi t^{N_0-1}} \sum_{n=0}^{N_2-1} \frac{
  c_{(n, \frac{4i}{3}, \frac{4i}{3})}}{2} 
F^{(1)}\left(t; \begin{array}{c}
N_0-n\\ \omega_2
\end{array} \right) +
\frac{c_{(n, -\frac{4i}{3}, -\frac{4i}{3})}}{2}
F^{(1)}\left(t; \begin{array}{c}
N_0-n\\ \omega_2^\star
\end{array} \right)\\
\displaystyle + \left(-\frac{1}{2i\pi}\right)^2 \frac{1}{t^{N_0-1}}
 \sum_{n=0}^{N_2-1} \left[ c_{(n, \frac{4i}{3}, \frac{4i}{3})}
F^{(2)}\left(t; \begin{array}{cc}
N_0-N_1+1 , & N_1-n \\
\omega_1 ,  & \omega_1
\end{array}\right)
\right.\\
\displaystyle +  c_{(n, \frac{4i}{3}, -\frac{4i}{3})}
F^{(2)}\left(t; \begin{array}{cc}
N_0-N_1+1 , & N_1-n \\
\omega_1 ,  & \omega_1^\star
\end{array}\right)
 +  c_{(n, -\frac{4i}{3}, \frac{4i}{3})}
F^{(2)}\left(t; \begin{array}{cc}
N_0-N_1+1 , & N_1-n \\
\omega_1^\star ,  & \omega_1
\end{array}\right) 
\\
\displaystyle + \left.  c_{(n, -\frac{4i}{3}, -\frac{4i}{3})}
F^{(2)}\left(t; \begin{array}{cc}
N_0-N_1+1 , & N_1-n \\
\omega_1^\star ,  & \omega_1^\star
\end{array}\right) \right]
+R(N_0, N_1, N_2, t).
\end{array}
\end{equation}
Note that, since we always use right-Borel-resummations, this induces the
following convention for the hyperterminants, when $\arg \sigma_j =
\arg \sigma_{j+1} \mod 2\pi$ for some $j$:
\begin{equation}\label{Hypertright}
F^{(l+1)}\left(t; \begin{array}{ccc}
M_0 , &\cdots , & M_l \\
\sigma_0 , &\cdots , & \sigma_l
\end{array}
\right)= \lim_{\epsilon \downarrow 0}
F^{(l+1)}\left(t; \begin{array}{cccc}
M_0 , & M_1,  &\cdots , & M_l \\
\sigma_0 e ^{-il\epsilon},& \sigma_1 e ^{-i(l-1)\epsilon} &\cdots , & \sigma_l
\end{array}
\right).
\end{equation}
For reasons explained in \cite{Old98}, the choices $N_0 =
3|\omega_1| |t[ = 12/3|t|$ and $N_1 = 2|\omega_1| |t[ = 8/3|t|$ and 
$N_1 = |\omega_1| |t[ = 4/3|t|$  give
optimal truncations, for which the remainder term behaves like 
$R(N_0, N_1, N_2, t) = \exp\big( -N_0 |t| \big) O(1)$, for 
$t  \in P_{\delta, \mu} (B)$. Numerical experiments are illustrated by
tables  \ref{table7} and \ref{table8}.

\begin{table}[thp]
\begin{center}
\begin{tabular}{|c|l|c|}
\hline
Optimal  $N$ & Estimates for $k_1$ &  Real error\\
\hline
$N_0 = 14$, $N_1 = 9$, $N_2 =5$  & -2.33810741077 & $0.31 \times 10^{-9}$ \\
\hline
\end{tabular}
\caption{Calculation for the first zero $k_1$ of the Airy function by
  level-2 hyperasymptotics.  
\label{table7}}
\end{center}
\end{table}

\begin{table}[thp]
\begin{center}
\begin{tabular}{|c|l|c|}
\hline
Optimal  $N$ & Estimates for $k_3$ &  Real error\\
\hline
$N_0 = 52$, $N_1 = 35$, $N_2 =17$  & -5.520559828095551059129855522 & $0.87 \times 10^{-26}$ \\
\hline
\end{tabular}
\caption{Calculation for the first zero $k_3$ of the Airy function by
  level-2 hyperasymptotics.  
\label{table8}}
\end{center}
\end{table}

\subsection{Higher level}\label{Hyper3}

From a theoretical viewpoint it is possible to perform higher
hyperasymptotics. This requires to go deeper in  the resurgent
structure of $X(t)$. The method is just like what we have done in \S \ref{ResForX}, even
if the alien calculus becomes somewhat complicated. For the
hyperasymptotics part, the existence of collinear singularities induce
the need of great care in choosing the correct branches of the
hyperterminants (in relation with (\ref{Hypertright})).  But,
according to the specialists, this problem can be mastered (see \cite{Old05}).

\end{document}